\documentstyle[10pt,epsfig]{article}
\begin{document}

\begin {center}
{\Large Re-analysis of data on $a_0(1450)$ and $a_0(980)$}

\vskip 5mm
{D.V.~Bugg\footnote{email address: D.Bugg@rl.ac.uk}},   \\
{Queen Mary, University of London, London E1\,4NS, UK}
\end {center}

\begin{abstract}
Four sets of data determining parameters of $a_0(1450)$ and $a_0(980)$
are re-analysed.
These are Crystal Barrel data for $\bar pp$ annihilation at rest to
$\eta \pi ^0\pi ^0$ in (i) liquid hydrogen and (ii) gas, and to (iii)
$K^0_L \pi ^\pm K^\mp$ and (iv) $\omega \pi ^+\pi ^- \pi ^0$ (mostly
$\omega \rho\pi$).
Dispersive corrections due to opening of inelastic thresholds
are treated fully.
This stabilises parameters of $a_0(1450)$ substantially.
The mass of its peak is $1448 \pm 13(stat) \pm 25(syst)$ MeV and its
mean full width at half maximum is $192 \pm 9 \pm 9$ MeV.
The pole position is
$M - i\Gamma /2 = 1432 \pm 13 \pm 25 - i(98 \pm 5 \pm 5)$  MeV.
At the peak, $\eta \pi$, $\omega \rho$ and $a_0(980)\sigma$ decay
intensities are in the ratios
$1 : 9.2 \pm 0.8 \pm 1.3 : 3.1 \pm 0.2 \pm 0.9$.
There is no evidence for a separate $a_0$ near 1300 MeV claimed by
Obelix.
Parameters of $a_0(980)$ are updated to $M = 987.4 \pm 1.0 \pm 3.0$
MeV, $g^2(\eta \pi) = 0.164 \pm 0.007 \pm 0.010$ GeV$^2$,
$g^2(KK)/g^2(\eta \pi) = 1.05 \pm 0.07 \pm 0.05$.
Its dominant second sheet pole in the $KK$ channel is at
$(989 \pm 1 \pm 5)-i(40 \pm 2 \pm 4)$ MeV.
Finally, the nature of the prominent  $J^{PC} = 0^{-+} \to \omega \rho$
signal in $\omega \rho \pi$ data is also clarified.


\end{abstract}

\section {Introduction}
The main objective of this work is to re-examine the parameters of
$a_0(1450)$.
It appears in the summary table of the Particle Data Group \cite
{PDG} although it has been observed decisively
in only one set of data, from Crystal Barrel on
$\bar pp \to \eta \pi ^0 \pi ^0$ at rest \cite {1450A}, \cite {1450B},
\cite {1450C}, \cite {1450D}.
There is further evidence from other data of the same
experiment.
It has been confirmed in the $\omega \rho$ channel in
$\omega \pi ^+\pi ^- \pi ^0$ data at rest \cite {rwpi}.
Its $KK$ decays are observed in $(K^0_L K^\pm )\pi ^\mp$ data at rest
\cite {Spanier}.
There is also evidence for it in $\bar pp \to (\eta '\pi ^0) \pi ^0$
\cite {1450E} and
$\bar pp \to (\eta \pi ^+\pi ^-\pi ^\pm )\pi ^\mp$ \cite {Nana}
at rest.

Its branching ratio to $\omega \rho$ in Ref. {\cite {rwpi}  is
a factor $\sim 11$ larger than to $\eta \pi$ (and will be revised here
slightly).
The fact that its branching ratio to $\eta \pi$ is $<10\%$ explains why
$a_0(1450)$ has been elusive in data for $\pi ^-p \to \pi \pi  n$.
An important point is that the  phase space for the
$\omega \rho$ channel has a rapid $s$-dependence, so it is
inappropriate to fit the $a_0(1450)$ with a Breit-Wigner amplitude of
constant width, as was done in the early work.
Its line-shape and the relation between magnitude and phase are
affected strongly by dispersive effects,
which are treated fully here.
Attention to this detail improves considerably the stability of fitted
parameters and makes the signal in both $\eta \pi$ and $\omega
\rho$ much clearer.

The $a_0(980)$ is examined along the same lines.
The dispersive effect due to the opening of the $KK$ channel plays a
critical role and again improves the quality of the fit to
$\eta \pi ^0 \pi ^0$ data.
Adler zeros are included into its $\eta \pi$ and $\eta '\pi$
decay channels.
For these reasons, its parameters change significantly from earlier
work.

The dispersive effects may be unfamiliar to experimentalists, though
well known to theorists since the 1950's.
Experimental analyses have conventionally been done with a
Breit-Wigner amplitude with denominator
\begin {equation} D(s) = M^2 - s - i \sum _j g^2_j\rho_j(s).
\end {equation}
Here $\rho_j(s)$ is the phase space for each decay channel
$j$ as a function of invariant mass squared $s$, possibly including a
form factor. The $g_j$ are coupling constants to each decay amplitude.
Let us write
\begin {equation}
D(s) = M^2 - s - \sum _j \Pi _j(s)
\end {equation}
with $\rm {Im}\, \Pi (s) = g^2_j \rho_j(s)$.
Because scattering amplitudes are analytic functions,  any
$s$-dependence of $\rm {Im}\, \Pi (s)$ necessarily leads to a
term in $D(s)$ given by
\begin{equation}
\rm {Re}\, \Pi_j(s) = \frac {1}{\pi}\rm {P}\, \int^\infty_{s(thr)}
\frac {\rm {Im}_j(s')ds'}{s' - s}.
\end {equation}
Here $\rm {P}$ denotes the Principal Value Integral and $s_{thr}$ is
the value of $s$ at threshold.
This is known as a dispersive contribution.
It is equivalent to evaluating loop diagrams.
If $\rho(s)$ changes rapidly, as it does at the opening of a sharp
threshold, the dispersive term becomes dominant and affects the
parameters of the resonance strongly.
Fig. 1 below illustrates the result for $a_0(980)$.
There is a prominent cusp in $\rm {Re}\,
\Pi_{KK}(s)$, centred at the threshold.
It plays a major role in locking the resonance to this threshold
\cite {sync}.
One objective of  the present work is to refine the parameters of
$a_0(980)$ to include this effect.

Consider next $a_0(1450)$.
The $\omega \rho$ threshold is quite sharp and has a large effect on
the line-shape near 1450 MeV.
There is a cusp at the $\omega \rho$ threshold which also acts as an
attractor.
This may be the reason that $a_0(1450) $ is higher in mass than
$f_0(1370)$ and $K_0(1430)$.

Section 2 reviews dispersive effects.
In principle they apply to all resonances.
Fortunately, resonances with broad thresholds may be approximated by
the pole term alone and this will be demonstrated here for $a_2(1320)$.
There may be small residual effects far from resonance, but in
practice these effects are tolerable at present.

Section 3 discusses fits to $\eta \pi ^0\pi ^0$ and
$\omega \pi ^+\pi ^- \pi ^0$ data, hence parameters of $a_0(1450)$.
In the present work, the widths of $a_0(1450)$ to $KK$ and $\eta \pi$
are small, so there is no longer significant overlap between these two
resonances and therefore little correlation between their parameters.
An incidental feature of the re-analysis of $\omega \pi ^+\pi ^-\pi
^0$ data is an improved understanding of the large
$J^{PC} = 0^{-+}$ $\omega \rho$ signal  observed there.

Section 4 gives results for $a_0(980)$ and Section 5 discusses $KK$
coupling of $a_0(1450)$ and parameters of $a_0(980)$.
The data on $K^0_L K^\pm \pi ^\mp$ do not give an accurate
determination of  their coupling to $KK$, but agree within sizable
errors with the better determination from $\eta \pi ^0\pi ^0$ data.
Section 6 summarises conclusions and makes some remarks on further
desirable work.

\section {Technicalities of the dispersive terms}
As an introduction, let us consider $a_0(980) \to KK$.
Mass differences between $K^+K^-$, $K^0_LK^\pm$ and $K^0\bar K^0$
will be ignored here because their separations are smaller than mass
resolution in data to be fitted.
There is a further reason.
The VES group has very recently presented data showing that the
$f_1(1285)$ decays to $3\pi$ \cite {VES}.
This violates isospin conservation and may well arise from mixing
between $a_0(980)$ and $f_0(980)$ due to mass differences in the
$KK$ thresholds.
Consideration of this problem requires a combined analysis with data
on $f_0(980)$.
It is necessary to take one step at a time and defer this for the
present, though one should bear in mind there may be some small
effect on parameters fitted to $a_0(980)$.

Ignoring mass differences, $\rho _{KK} = \sqrt {1 - 4m^2_K/s}$,
where $m_K$ is the mean kaon mass, 495.663 MeV.
As $s \to \infty$, the phase space factor $\to 1$.
Without any form factor, the dispersion integral of Eq. (3) diverges.
Therefore a form factor
 \begin {equation}
 F_{KK} = \exp (-\alpha k^2)
 \end {equation}
is used to multiply $g_{KK}$.
Here $k$ is the momentum of each kaon in the $KK$ rest frame.
This well known form factor assumes a Gaussian source with RMS
radius $R$ given by $\alpha = R^2/6$.
It turns out that the same value of $\alpha$ succeeds in fitting all
resonances and avoids a multiplicity of parameters.
It optimises at $\alpha = (2.0 \pm 0.25)$ (GeV/c)$^{-2}$, corresponding
to $R = 0.68 \pm 0.04$ fm.

\begin {figure}  [htb]
\begin {center}
\vskip -13mm
\epsfig{file=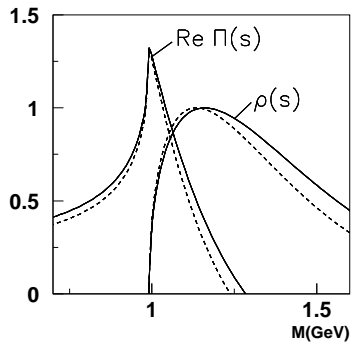,width=6.5cm}\
\vskip -6mm
\caption{$KK$ phase space $\rho '(s)$, normalised to 1 at
its peak and $\rm {Re}\, \Pi (s)$ (full curves). Dashed curves show
results for $\alpha = 2.5$ (GeV/c)$^{-2}$.}
\end {center}
\end {figure}

Library subroutines are available for evaluating the Principal Value
Integral.
Results are shown in Fig. 1 for the product
$\rho ' = \rho_{KK}(s)F^2_{KK}(s)$ and for $\rm {Re}\, \Pi(s)$ .
For display purposes, the normalisation is chosen so that $\rho'$
including the form factor peaks at 1.
There is a large cusp in the real part, somewhat larger than the peak of
$\rho '$.
The dashed curves show results with $\alpha = 2.5$ (GeV/c)$^{-2}$.
In the mass range where the $a_0(980)$ is strong, the
sensitivity to $\alpha$ is quite small; it comes into play only in the
wings of the resonance, where other resonances may mask its effects.

There are two practical points concerning $\rm {Re}\, \Pi (s)$.
Although it is responsible for attracting $a_0(980)$ to the $KK$
threshold, it is convenient to make a subtraction in the Breit-Wigner
denominator on resonance:
 \begin {equation}
 D(s) = M^2 - s - \sum _j\rm {Re}\, [\Pi _j(s) - \Pi _j(M^2)] -
i g^2_j \rho '(s).
 \end {equation}
Secondly, it is convenient to evaluate the dispersion integral as a
2-dimensional array against $s$ and $\alpha$.
A simple sub-routine interpolates in this table.
When fitting data, one can then optimise $\alpha$, $M^2$ and $g^2$
with a standard optimisation program such as Minuit.
The dispersive term is proportional to $g^2$.
Just below the $KK$ threshold, it varies as
$g^2_{KK}(4m^2_K - s)/s$  (see algebra in Ref.
\cite {sync}, Eq. 15).
This resembles the term $(M^2 - s)$ in $D(s)$.
Consequently $M$ and $g^2$ become strongly correlated
unless there are data determining $g^2$ separately for every channel.

Because of the correlations, the
convergence of the fit it rather poorer than for a simple
Breit-Wigner resonance of constant width but still adequate.
It is in fact better to let the programme optimise the parameters.
The alternative, a grid search over $M$ and $g^2$, is subject to
the correlations between them.
Standard optimisation programmes work with eigenvectors and
circumvent the correlations.

A general point is that all resonances are subject to opening
thresholds, hence dispersive effects.
However, it fortunately turns out that for broad thresholds
the net effect of the dispersive terms becomes small within one full
width of the pole.
The data can then be parametrised directly in terms of the pole term
$\propto 1/(s - s_{pole})$.
This conclusion emerged from work on $f_0(1370)$ concerning
$\sigma \sigma $ and $\rho \rho$ thresholds \cite {f01370}.
For broad thresholds, the dispersive terms have significant effects
only far from the pole, and may not be trustworthy there because of
uncertainties in form factors.
In the present work, thresholds for
$a_1(1260) \to (\rho \pi )_{L=1}$,
$a_2(1320) \to (\rho \pi )_{L=2}$, $(\eta \pi )_{L=2}$ and $(KK)_{L=2}$
have been treated fully using the dispersive term.
These thresholds open fairly gently because of the centrifugal barriers
for orbital angular momentum $L$ in the decays.
The conclusion is the same as in \cite {f01370}: the line-shapes of
these resonances are affected rather little except for their tails.

However, in fitting $\omega \rho \pi$ data, the rather sharp
$\omega \rho$ threshold does affect the fitted resonances quite
strongly.
The cusp at the $\omega \rho$ threshold is broadened by the line-shape
of the $\rho$.
This line-shape may be included in the evaluation of $\omega \rho$
phase space, then the dispersive effect can be evaluated from the
phase space.
Suppose as an example $a_0(1450) \to \omega \rho$, followed by
$\rho \to \pi \pi$.
The 3-body phase space for $\omega \rho$ is given by the
integral
\begin {equation}
\rho '_{\omega \rho}(s) = \int ^{(\sqrt {s} - m_\omega)^2} _{4m^2_\pi}
\frac {ds_1}{\pi} \frac {4|k||k_1|}{\sqrt {s s_1}}
|T_\rho (s_1)|^2 \exp (-2\alpha k^2),
\end {equation}
where $T$ is the Breit-Wigner amplitude for the $\rho$.
Also $s$ refers to the $a_0(1450)$ and $k$ to the momentum of the
$\omega$ or $\rho$ in the $a_0$ rest frame; $s_1$ and $k_1$ refer to
the $\rho$ and the momenta of the pions in its rest frame.
When there is angular momentum in the decay to $\rho \omega$,
a centrifugal barrier needs to be included.

\section {Fits to $\bar pp \to \eta \pi ^0 \pi ^0$ and
$\omega \pi ^+\pi ^-\pi ^0$}
The $a_0(1450)$ was discovered in Crystal Barrel data for
$\bar pp \to \eta \pi ^0 \pi ^0$ at rest \cite {1450A}, \cite {1450B},
\cite {1450C}, \cite {1450D}.
It also appears in the $\omega \rho$ channel in $\omega \rho \pi$
data at rest \cite {rwpi}, and in the $KK$ channel in
$K^0_L K^\pm \pi ^\mp$ data at rest \cite {Spanier}.
The latter will be discussed in Section 5, but it turns out that the
systematic error in its coupling to $KK$ is rather large.
Its $KK$ coupling is consistent with the SU(3) prediction, and will be
fixed to that value.
The same applies to the $\eta '\pi$ coupling.
An analysis of data on $\bar pp \to \eta '\pi ^0\pi ^0$ at rest gave
results consistent with this prediction \cite {1450E}.
The effects of both $KK$ and $\eta '\pi$ channels on the line-shape
of $a_0(1450)$ are similar to $\eta \pi$ and quite small.

A preliminary comment is required on the fit to data for
$\bar pp \to \omega \pi ^+\pi ^- \pi ^0$, discussed in
sub-section 3.9.
The earlier publication did include dispersive effects.
The fit to these data changes rather little in the combined fit
with $\eta \pi ^0 \pi ^0$ data.
The main improvement to parameters of $a_0(1450)$ comes from
the $\eta \pi ^0 \pi ^0$ data.

The decays of consequence for the line-shape are $\eta \pi$,
$\omega\rho$ and $a_0(980)\sigma$.
If $\omega\rho$ is the only strong decay channel, it restricts
the maximum possible full-width severely.
This is because the rapid increase in $\rho _{\omega \rho}$ with $s$
inflates the Breit-Wigner denominator and cuts off the upper side of
the resonance.
The consequence in Ref. \cite {rwpi} was that the optimum fitted
full-width was only 110 MeV.

This width is too small for good consistency with the
$\eta \pi ^0 \pi ^0$ data.
Some other broad threshold is needed for an acceptable fit.
This is provided by the $a_0(980)\sigma$ decay.
Data for $\bar pp \to \eta \pi ^+\pi ^-\pi ^-\pi ^-$ at rest
\cite {Nana} were found to contain some $a_0(1450)$ signal in
$\eta \pi^+\pi ^-\pi ^+$,
improving log likelihood by 32 for 2 fitted parameters;
this is statistically $>7$ standard deviations.
However, there was no optimum when the mass and width of $a_0(1450)$
were scanned.
The branching fraction for the $\eta 4\pi$ final state is a factor 14
larger than for $\eta \pi \pi$, with the result that the allowed
branching fraction of $a_0(1450) \to a_0(980)\sigma$ could be as
much as 4.3 times that of $\eta \pi$.
It now turns out that including the $a_0(980)\sigma$ threshold supplies
the required broad component in $a_0(1450)$ decays and improves
markedly the fit reported here.
The required branching ratio to $\eta 4\pi$ is only
slightly smaller than that fitted in Ref. \cite {Nana}, so it appears
to be a genuine signal.

Ideally the $\eta \pi ^+\pi ^-\pi ^+\pi ^-$ data should be
included in the present fits.
Unfortunately those data have been lost, so this is not possible
without major work reprocessing them from raw data.
This is not worth the effort, since they did not constrain
the mass and width of $a_0(1450)$.
All that matters is the magnitude of the fitted signal and the
upper limit on the $a_0(980)\sigma$ branching fraction; these
can be taken from the earlier publication.

\begin {figure}  [htb]
\begin {center}
\vskip -16mm
\epsfig{file=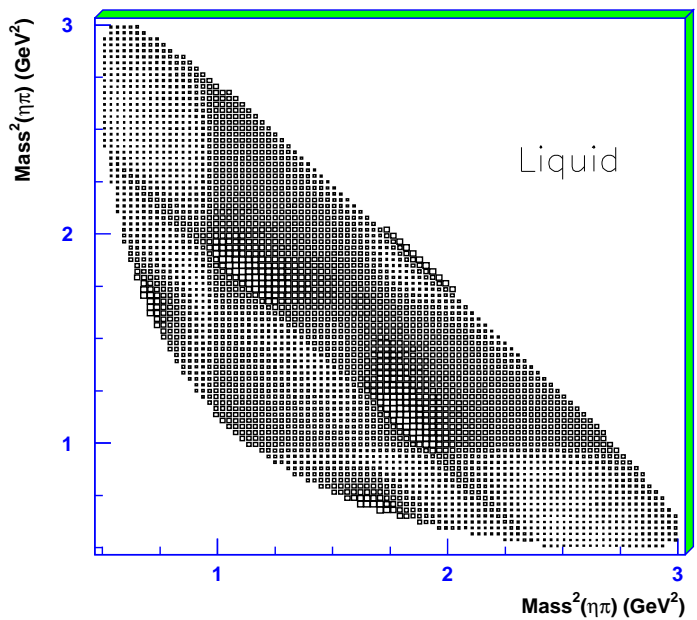,width=10.0cm}\
\vskip -7mm
\caption{The Dalitz plot for $\bar pp \to \eta \pi^0\pi ^0$ at rest
in liquid hydrogen.}
\end {center}
\end {figure}

\subsection {Features of the $\eta \pi ^0\pi ^0$ data}

The Dalitz plots for data in liquid and gaseous hydrogen are shown
in binned form in Figs. 2 and 3.
There are $\sim 280,000$ events in liquid hydrogen with experimental
background $<1\%$.
A minor detail is that any bins overlapping the edges of the Dalitz
plots have been removed from Figs. 2 and 3 and the fits.
There are also some further bins immediately adjoining edge bins
and showing questionable behaviour.
This can arise if an event lies outside the true Dalitz plot
before the kinematic fit.
That fit enforces the constraints of energy-momentum
conservation and the masses of $\pi ^0$, $\eta$ and $\omega$.
It pulls events inside the Dalitz plot, but there is some tendency for
them to congregate towards the edges.
These bins are easily identified and removed because the fit is
systematically lower than data.
A total of 18 out of 3582 bins are removed for this reason, though
effects on fitted parameters are tiny.

Statistics for $\eta \pi \pi$ are so high ($\sim 280,000$ events) that
it was not possible to equal those statistics in the Monte Carlo
simulation. (Only a few per-cent of events survive the data
selection).
It is assumed that the acceptance is uniform, in accord with
observations for present data, $\bar pp \to \eta \eta \pi ^0$ and $3\pi
^0$.
The final fit has a $\chi^2$ of 2.9 per bin.
A similar value was obtained in fitting $\bar pp \to 3\pi ^0$
\cite {f01370}.
Examination of the present fit reveals no systematic deviation
across the Dalitz plot associated with fitted components.
There are possible slow variations with $\chi^2 $ up to 10 which could
be associated with small systematic effects in the slowly varying $\pi
\pi$ S-wave or alternatively could arise from small variations in
experimental acceptance.
Any departure from uniform acceptance over the width of $a_0(1450)$
has an effect much smaller than errors.
However, it has been necessary to scale statistical errors
to account for the mean $\chi^2$ per bin.

\begin {figure}  [htb]
\begin {center}
\vskip -17mm
\epsfig{file=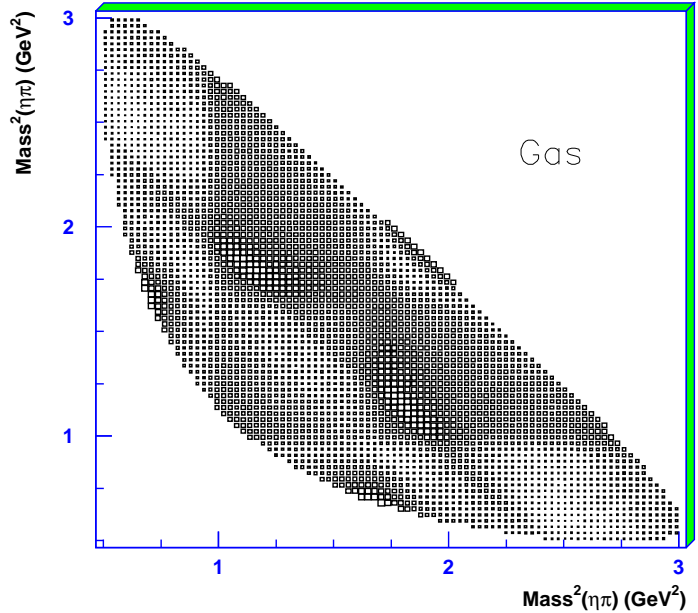,width=10.0cm}\
\vskip -7mm
\caption{The Dalitz plot for $\bar pp \to \eta \pi^0\pi ^0$ at rest
in gaseous hydrogen.}
\end {center}
\end {figure}

Figs. 2 and 3 resemble one another closely, showing that the effects of
P-state annihilation are small.
Both sets of data are fitted fully, and the final fit gives $7.4\%$
P-state contribution, in close agreement with an earlier determination
\cite {BRs}.
This come mostly from $^3P_1$ and $^3P_2 \to  a_2(1320)\pi$
and $a_2(1700)\pi$, and $^3P_1 \to \eta \sigma$ and $\eta f_0(980)$.
The former plays an important role in fitting the angular dependence
of the prominent $a_2(1320)$ bands.
The effect of the latter two components is visible along the $f_0(980)$
band, where interferences between $f_0(980)$ and $\sigma$ affect the
apparent width of the $f_0(980)$ in the data.
Ultimately P-state annihilation has little effect on
fitted parameters of either $a_0(980)$ or $a_0(1450)$.
P-state production of $a_0(1450)$ is inhibited by a centrifugal barrier
and makes only a very weak contribution $(0.27\%)$.

The two $a_2(1320)$ bands interfere constructively at the upper
right-hand edge of the Dalitz plots.
Interference between the two $a_2$'s builds a bridge between them along
this edge.
The bands appear to be not quite vertical/horizontal.
In the analyses of the 1990's, this deviation was fitted by a broad
$\eta \pi$ P-wave resonance with ill-defined mass and a large width
of $\sim 600$ MeV.
Those parameters are inconsistent with what is now known about the
$\eta \pi$ P-wave.
The current fits are made with the $\pi _1(1400)$ parameters fitted to
Crystal Barrel data on $\bar p n \to \pi ^-\pi ^0 \eta$
\cite {Wolfgang}.
In those data, there is a significant P-state contribution
because the process $^1P_1 \to \pi \pi_1(1400)$ goes via the S-wave.
In  present $\eta \pi ^0\pi ^0$ data, there is now a small (0.9\%)
P-state contribution from $\pi (1400)$ and only $ 0.6\%$ in S-state
annihilation.

Another distinctive feature of the Dalitz plots is
a sharp `edge' in $\eta \pi$ coinciding accurately with the $KK$
threshold.
This is due to the the opening of the $KK$ threshold for $a_0(980)$.
At this threshold, $\rho _{KK}$ changes from real to imaginary as one
crosses the threshold from above to below.
The amplitude for $a_0(980)$ therefore turns
in phase by $90^\circ$.
Consequently, interference with the $\pi \pi$ S-wave
changes dramatically.
The precise form of the `edge' is therefore sensitive
to the relative coupling of $a_0(980)$ between $\eta \pi$ and $KK$.

\begin {figure}  [htb]
\begin {center}
\vskip -17mm
\epsfig{file=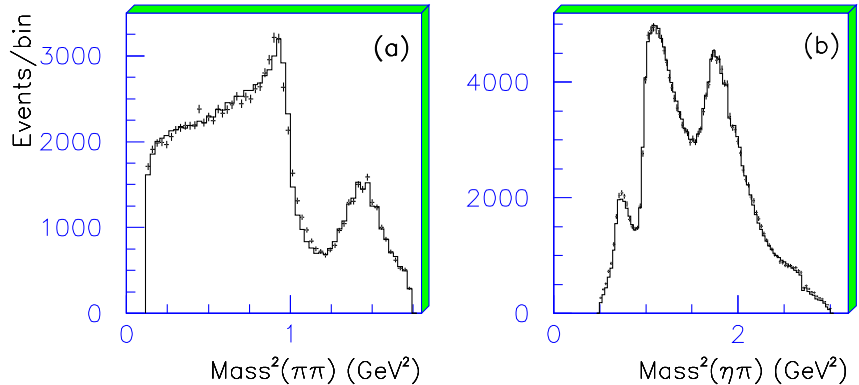,width=12.0cm}\
\vskip -7mm
\caption{Mass projections for (a) $\pi \pi $ and (b) $\eta \pi $
for $\bar pp \to \eta \pi ^0\pi ^0$ at rest.
Points with errors show the data; histograms show the fit.}
\end {center}
\end {figure}

Fig. 4 shows mass projections for $\pi ^0\pi ^0$ and $\eta \pi ^0$  in
liquid hydrogen and  the fit.
Fig. 4(b) is the easier to understand.
The first (left-hand) peak is a reflection of the
$a_2(1320)$ at the left-hand side of the Dalitz plot.
The sharp rise to the second peak is caused by the `edge' due to
$a_0(980)$ and its interferences with $a_2(1320)$.
The third peak is directly due to $a_2(1320)$.
The sudden drop at high mass is due to the $a_2(980)$ `edge' crossing
the right-hand side of the Dalitz plot.
The quality of the data (and fit) illustrate the information available
on $a_0(980)$ and its coupling to $KK$.
Note that the $a_0(1450)$ is not directly visible in Fig. 4(b).

In Fig. 4(a), there is one high point at $s_{\pi \pi} = 0.44$ GeV$^2$.
It does not correlate with anything and appears to be a statistical
storm.
The first peak to its right is  again due to $a_0(980)$
and its interferences with $a_2(1320)$ and the $\pi \pi$ S-wave.
The second peak is a reflection of $a_2(1320)$ on the lower side of
the Dalitz plot.

The cusp in $a_0(980)$ at the $KK$ threshold is sufficiently narrow
that it is necessary to fold in the mass resolution for bins
adjoining the KK threshold.
The mass resolution is a Gaussian with a $\sigma$ of 9.5 MeV.
This number is derived from data on $\bar pp \to \pi ^0\eta' (958)$
\cite {resln}, where the fitted width of the $\eta '$ is readily
measured.
The folding is done using Gaussian 12 point integration over the
bins concerned.

\subsection {The $\pi \pi$ S-wave amplitude}
This is the third major component in the $\eta \pi ^0\pi ^0$ data.
Since the earliest publications in the 1990's, our knowledge of the
$\pi \pi$ S-wave amplitude has improved greatly.
Today, the $\sigma$ pole is well known from (a) the BES2 data on
$J/\Psi \to \omega \pi ^+\pi^-$, where it produces a strong peak
at $\sim 500$ MeV \cite {WPP}, (b) the calculations of Caprini,
Colangelo and Leutwyler using the Roy equations to constrain the
$s$-dependence of the elastic amplitude \cite {Caprini}.

The $\pi \pi$ elastic scattering amplitude may be written in the
form
\begin {equation}
f_{el}(s) = N(s)/D(s),
 \end {equation}
where $N(s)$ is real and must be equal to $-\rm {Im}\, D(s)$
below the $KK$ threshold.
In a production reaction, $D(s)$ must be the same for the
$\sigma$ pole as in elastic scattering (Watson's theorem
\cite {Watson}).
However, $N(s)$ is allowed to be quite different between production and
elastic scattering \cite {Extended}.
The strong peak close to 500 MeV in BES data for
$J/\Psi \to \omega \pi ^+\pi ^-$ is fitted accurately taking $N(s)$ to
be constant.
There is then accurate agreement \cite {sigpole} between the pole
observed in these data and the elastic phase shifts predicted by
Caprini et al.

The recent fits to data on $\bar pp \to 3\pi ^0$ \cite {f01370}
require a 2-component form for the S-wave production amplitude:
\begin {equation}
f_{prodn} = \Lambda _1 f_{el}(s) + \Lambda _2/D(s),
\end {equation}
where $\Lambda_{1,2}$ are complex coupling constants: i.e. a
coherent sum of the elastic amplitude and the pole term.
This 2-component prescription also fits the $\eta \pi ^0\pi^0$ data,
with different $\Lambda$ to those for $\bar pp \to 3\pi ^0$.
This prescription  will play an essential role throughout the present
work, including the fit to data on $\bar pp \to K^0_L K^\pm \pi ^\mp$
where both the $\kappa$ pole and the elastic $K\pi$ amplitude
contribute.
Similar variations of the $\pi \pi$ S-wave amplitude are well known
in decays of $\Upsilon '$ to $\Upsilon \pi \pi$ and are discussed
in detail by Simonov and Veselov \cite {Simonov}.

Two alternative prescriptions are available for the $\pi \pi $ S-wave,
from Refs. \cite {sigpole} and \cite {f01370}.
The latter is fitted to data for $\bar pp \to 3\pi ^0$, where the
$\pi\pi$ mass range extends to 1.74 GeV.
For $\bar pp \to \eta \pi ^0 \pi ^0$, the  $\pi \pi$ mass range stops
at 1.329 GeV.
The $4\pi$ inelasticity is quite small up to this mass.
The two alternatives lead to only minor differences in the quality of
fit to $\bar pp \to \eta \pi ^0\pi ^0$.
The first prescription is simpler and faster and is used for final
fits.

A question arises whether to assume the $a_0(980)\sigma$ channel
is produced via the $\sigma$ pole or the elastic $\pi\pi$
amplitude.
If the latter is used, $a_0(980)\sigma$ phase space rises too slowly
to have much effect over the mass range of $a_0(1450)$.
Some production of the $\pi \pi$ S-wave via its
pole term is needed and is what is used here.
It is also what was fitted to $\eta 4\pi$ data.
The mean mass of the $\sigma$ is then $\sim 470$ MeV and the full
width is $\sim 500$ MeV.
The $a_0\sigma$ phase space is then substantial at 1450 MeV.

\begin {figure}  [htb]
\begin {center}
\vskip -18mm
\epsfig{file=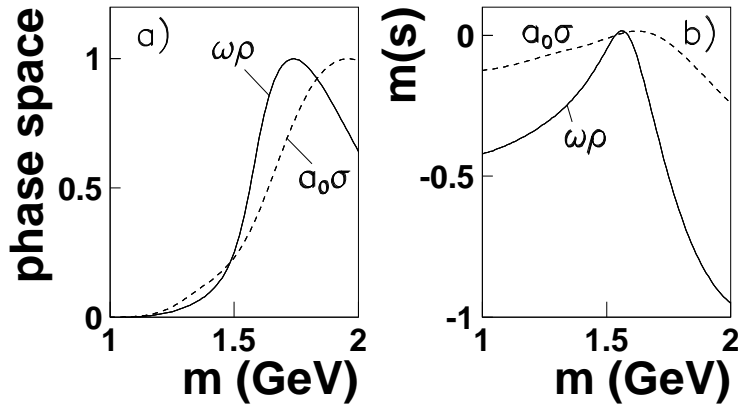,width=11.0cm}\
\vskip -4mm
\caption{(a) Phase space for $\omega \rho$ (full curve) and
$a_0(980)\sigma$ (dashed); (b) contributions to $\rm {Re}\, \Pi (s)$ for
$a_0(1450)$.}
\end {center}
\end {figure}

Fig. 5(a) shows $\omega\rho$ and $a_0(980)\sigma$ phase space.
They both peak in the mass range 1.65--2.0 GeV because of the form
factor.
Fig. 5(b) shows the subtracted form $m(s) = \rm {Re}\, \Pi(s) - \rm
{Re}\, \Pi (M^2)$ for $a_0(1450)$ with the normalisation of final fits.
To a first approximation, they are proportional to  the gradient of
phase space.
Their sum is roughly half the magnitude of $M^2 - s$, so their effects
on the line-shape of the resonance are quite large.

\subsection {Treatment of $f_0(980)$}
A further element required to fit the $\eta \pi ^0\pi ^0$ data is from
$f_0(980)$.
A full re-analysis of $f_0(980)$ parameters including the dispersive
effect is a major undertaking requiring fits to the many sets of data
in which it is prominent.
For present data, the line-shape of $f_0(980)$ is not critical.
Fig. 6(b) below will illustrate the blurring of the threshold
cusp in $a_0(980)$ due to mass resolution.
The blurring is even more severe for $f_0(980)$, which has a full-width
at half-maximum of only $34 \pm 8$ MeV \cite {phipp}, compared with the
mass resolution of $\pm 9.5$ MeV.
The mass resolution is folded with the line-shape of $f_0(980)$ in the
fit to data.
The effect of the  threshold cusp is a marginal decrease in the width
compared with the BES parametrisation.
This small perturbation has negligible effect on the parameters fitted
to $a_0(980)$ and $a_0(1450)$, because information on these two
resonances comes from regions of the Dalitz plot having only modest
overlap with $f_0(980)$.

\subsection {Fits to $a_0(1450)$}
Fits have been made simultaneously to the four set of data listed in the
Abstract.
The precise formula fitted to $a_0(1450)$ needs discussion.
A form factor is needed in calculating dispersive terms, in order to
make the dispersion integrals converge.
However, as Fig. 5(a) shows, the form factor plays a strong role only
above 1650 MeV, well above the $a_0(1450)$.
The form factor is therefore an unnecessary elaboration over the
mass range covered by $a_0(1450)$.
For simplicity, it is therefore dropped in the amplitude fitted to data.

The $a_0(1450)$ amplitude may be written
\begin {equation}
f(1450) = 1/[M^2 - s - (\Pi (s) - \Pi(M^2)) - i \sum_j g^2_j\rho _j(s)],
\end {equation}
where the sum runs over $\eta \pi$, $KK$, $\eta '\pi$, $\omega \rho$ and
$a_0(980)\sigma$ channels.
A slightly rearranged formula will be given later in the light of
observed results.
Values of $g^2$ for $KK$ and $\eta '\pi$ at the peak (i.e. near the
pole) will be fixed to SU(3) predictions, which depend on the angle
$\phi = 54.7^\circ - \Theta_{PS}$, where $\Theta _{PS}$ is the
pseudoscalar mixing angle.
Values of $\phi$ may be obtained from analysis of radiative decays of
vector (V) and pseudoscalar mesons (P) mesons.
Escribano and Nadal analyse all existing data and conclude there is no
significant evidence for a gluonic component in $\eta$ or $\eta '$
\cite {Nadal}.
It seems prudent to use results without that component.
They then find $\phi = (41.5 \pm 1.2)^\circ$.
Thomas does a similar analysis with an identical conclusion
\cite {Thomas}.
Data on $J/\Psi \to VP$ decays also give a less precise result:
$\phi = (40.5 \pm 2.4)^\circ$ \cite {Escribano}.
The weighted mean $\phi = (41.3 \pm 1.2)^\circ$ will be used here.
Then
\begin {eqnarray}
g^2_{\eta '\pi}/g^2_{\eta \pi} &=& \tan^2 \phi = 0.772 \pm 0.068, \\
g^2_{KK}/g^2_{\eta \pi} &=& 1/(2\cos ^2 \phi ) = 0.886 \pm 0.034.
\end {eqnarray}

\subsection {Treatment of branching ratios}
The relative value $g^2_{\omega \rho }/g^2_{\eta \pi}$ is obtained
from relative branching fractions of $a_0(1450)$ in $\eta \pi \pi$ and
$\omega \rho \pi$ data.
However there are two points which need to be taken into account.

Firstly, the observed branching fractions for each resonance in $\bar
pp$ data must be obtained by folding the phase space factors for each
channel with the line-shape of the resonance, using integrals of the
form
\begin {equation}
 I_j = \int \frac {ds\, g^2_j(s)\rho_j(s)\, k'}{|D(s)|^2}.
\end {equation}
The factor $k'$ is the momentum with which the resonance is produced in
$\bar pp \to \pi + a_0(1450)$; it allows for the phase space
corresponding to the length of the $a_0(1450)$ band as a function of
$s$ on the Dalitz plot.

Secondly, there is an important point of principle concerning how to
account for interferences.
Data are fitted including all the interferences, not only between
different resonances but also including, for example, two $a_2(1320)$
appearing in $\eta \pi ^0 \pi ^0$ data.
The coupling constants $\Lambda$ are determined by the fit;
but then, for use in Eq. (12), intensities of individual components
must be evaluated from these $\Lambda$ without the interferences.
The two $a_2$'s contribute 30.3\% of $\eta \pi ^0\pi ^0$ data, but
5.8\% of this arises from interference between the two bands.

There are even larger effects for $a_0(1450)$.
In $\eta \pi \pi$, there are constructive interferences between the two
$a_0(1450)$.
Including interferences, they contribute $5.44\%$ of the cross
section, but without interferences, this drops to $A = 3.48\%$.
In $\omega \pi ^+\pi ^-\pi ^0$ data, there are three charge states
for $a_0$ in the amplitude $(a_0^+\pi ^- - a_0^0\pi ^0 +a_0^-\pi^+)$,
where signs arise from isospin Clebsch-Gordon coefficients.
There are therefore some destructive interferences.
With this interference included, the $a_0$'s contribute
$3.49\%$ of $\omega \rho \pi$ data, but without them $B=4.86\%$.
The ratio $B/A$ determines $3g^2_{\omega \rho}I_{\omega \rho}/
g^2_{\eta \pi}I_{\eta \pi}$.
For the $a_0(980)$, interference effects are quite small, because
the peak of the resonance is narrow.

It is necessary to arrange, iteratively, that the fitted branching
ratio between $\eta \pi$ and $\omega\rho$ signals is consistent with
the fitted value of $g^2_{ \omega\rho}/g^2_{\eta \pi}$.
Table 1 lists the percentages of the signals
fitted to $\eta \pi ^0 \pi ^0$ data including interferences.
These do not add up to $100\%$ because of interferences.
\vskip -4mm
\begin{table}[htb]
\begin {center}
\begin{tabular}{cc}
\hline signal & Percentage  \\\hline
$\pi \pi $ S-wave  & $9.4  \pm 0.3 $ \\
$f_0(980)$        & $11.7 \pm 0.2 $ \\
$a_0(980)$         & $12.8 \pm 0.2 $ \\
$a_0(1450)$        & $5.44 \pm 0.25$ \\
$a_2(1320)$        & $30.3 \pm 0.2$ \\
$a_2(1700)$        & $6.2 \pm 0.2  $\\
$\pi_1(1405)$      & $0.6 \pm 0.2$ \\
P-states           & $7.4 \pm 0.5$ \\\hline
\end{tabular}
\caption{Intensities of signals fitted to $\bar pp \to \eta \pi ^0\pi
^0$ data.}
\end{center}
\end{table}

\subsection {Comments on the fit to $a_0(1450)$}
The fit to $a_0(1450)$ improves significantly compared with work in the
1990's where a Breit-Wigner resonance of constant width was assumed.
Parameters of $a_0(980)$ and $a_0(1450)$ were correlated significantly
in that early work, because the large and constant width of
$a_0(1450)$ made it overlap $a_0(980)$ significantly.
The mass of $a_0(1450)$ could move between 1450 and 1510 MeV as
$\Gamma$ was varied.
Now the width of $a_0(1450)$ near 1 GeV is restricted to $\eta \pi$ and
$KK$ and the width to $\eta \pi$ is only 19 MeV at a mass of 1 GeV.
The result is that $a_0(980)$ and $a_0(1450)$ are now almost
uncorrelated.
With the $s$-dependent forms used here, the peak position is very
stable in the range 1440-1460 MeV, with an optimum at 1448 MeV.

Furthermore, the fitted $a_0(1450) \to \eta \pi$ signal increases from
3.0\% to 5.4\%.
The data clearly prefer the $s$-dependent form.
The full width of $a_0(1450)$ at half maximum decreases substantially
from the $265 \pm 13$ MeV quoted by the Particle Data Group \cite {PDG}
to $192 \pm 9(stat) \pm 9(syst)$ MeV.
This is inevitable in view of the rapidly increasing $a_0(980)\sigma$
and $\omega \rho$ signals, which make the Breit-Wigner denominator
cut off the line-shape at high mass.
The PDG value is subject to serious systematic error from the
assumption of constant width.

\begin {figure}  [htb]
\begin {center}
\vskip -16mm
\epsfig{file=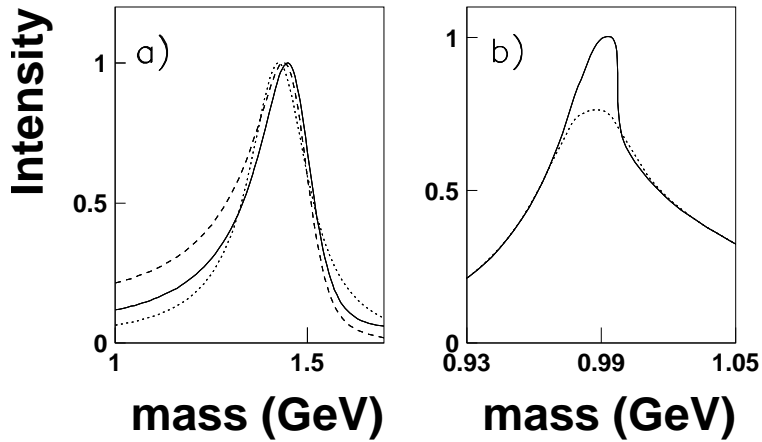,width=11.0cm}\
\vskip -3mm
\caption{(a) Natural line-shape of $a_0(1450)$ (full curve) and with
the effect of $\bar pp$ phase space included (dashed curve);
the dotted curve shows a Breit-Wigner resonance of constant width
agreeing at half-height with the full curve.
(b) Natural line-shape of $a_0(980)$ (full curve) and including the
mass resolution of the Crystal Barrel detector (dotted).}
\end {center}
\end {figure}
The $a_0(1450)$ line-shape for an isolated resonance, i.e. without the
factor $k'$ of Eq. (12), is shown by the full curve of Fig. 6(a).
What is plotted is $|1/D(s)|^2$, i.e. ignoring any phase space effects
in the numerator of the amplitude.
The line-shape observed in $\bar pp$ data including the factor $k'$ is
shown by the dashed curve.
A Breit-Wigner line-shape with constant width is shown by the dotted
curve, agreeing at half-height with the full curve.
The true  line-shape is asymmetric because of the rising
phase space for $\omega  \rho$ and $a_0(980)\sigma$ and also because of
the dispersive term $\Pi (s)$ in the Breit-Wigner denominator.

Fig. 6(b) shows the line-shape of $a_0(980)$ without and with the effect
of mass resolution of the Crystal Barrel detector.
The $a_0(980)$ is cut almost in half by the opening of the $KK$
threshold.
At this threshold, the line-shape drops rapidly because of the
$KK$ width in the Breit-Wigner denominator.
Many theorists base calculations on the 50--100 MeV width of the
$a_0(980)$ quoted by the PDG.
This is the full width at half-maximum.
The values of $g^2_{\eta \pi }$ and $g^2_{KK}$ are both $\sim 160$ MeV,
comparable with other resonances.

\begin {figure}  [htb]
\begin {center}
\vskip -12mm
\epsfig{file=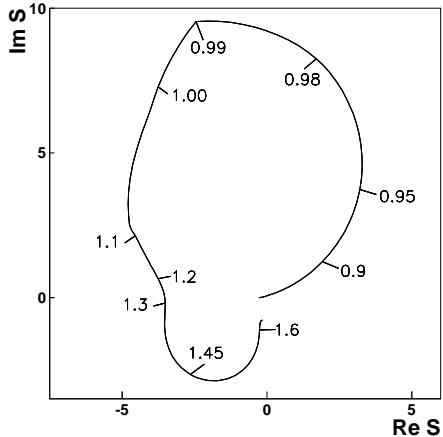,width=6.5cm}\
\vskip -3mm
\caption{Argand diagram for the coherent sum of $a_0(980)$ and
$a_0(1450)$, excluding the effect of experimental mass resolution.
Masses are shown in GeV.}
\end {center}
\end {figure}
The Argand diagram for the coherent sum of $a_0(1450)$ and
$a_0(980)$ is shown in Fig. 7, excluding the effect of mass resolution.
The maximum amplitude for $a_0(1450)$ is at 1448 MeV, where the
phase of the $\eta \pi$ amplitude is only $\sim 50^\circ$.
The phase goes through $90^\circ$ only at 1536 MeV.
This is the mass $M$ in the Breit-Wigner denominator.
The full curve of Fig. 8 shows the Argand diagram of $a_0(1450)$ drawn
from $1/D(s)$ alone.
It appears to lie on its side because it is cut off at high mass by
the rapid increase of $a_0\sigma$ and $\omega \rho$ phase space.
The value of $M$ in the Breit-Wigner denominator is a derived quantity,
rather strongly dependent on $g^2(\omega \rho)$ and
$g^2(a_0(980)\sigma)$ and their form factors.
Accordingly, $M$ has a factor 2 larger error than the peak mass, which
responds directly to the magnitude of the $\eta \pi$ amplitude.

An important check is whether $a_0(1450)$ really requires a resonant
loop like that shown on Fig. 7.
The first check is to replace the resonant form with its absolute
magnitude, deleting its phase variation.
As expected, $\chi ^2$ increases by 297.3 (after renormalising to allow
for the fact that $\chi ^2$ is 2.9 per data point); this is  a 17
standard deviation effect.

A more delicate check is to break the mass range from 1315 to 1675 MeV
into 30 MeV bins and optimise the $a_0(1450)$ signal in each bin.
The result is compared with the Argand diagram of $a_0(1450)$  alone on
Fig. 8.
The individual bins follow the expected loop closely up to 1540 MeV;
dotted lines show the movement of each bin from the analytic formula.
Up to this mass, there is large interference between $a_2(1320)$ and
$a_0(1450)$, providing strong constraints on its phase variation with
mass.
Above this, three of the four remaining points show a rather large
scatter.
Above 1560, the effect of $a_2(1700)$ becomes more important than
that of $a_2(1320)$.
Final fits use Crystal Barrel parameters for $a_2(1700)$:
$M = 1660$ MeV, $\Gamma = 280$ MeV \cite {Bochum}.
The main problem is that the corner of the Dalitz plot above
$m_{\eta \pi} = 1560$ MeV is a cramped area in which to separate spin
zero components from spin 2.
The $a_2(1700)$ has significant contributions from all three initial
states $^1S_0$, $^3P_1$ and $^3P_2$.
These allow it to simulate a spin 0 contribution to some
extent, despite the existence of the data in gas which help
determine P-state contributions.
The most likely explanation of the discrepancies above 1560 MeV
is a poor separation between $a_2(1700)$ and $a_0(1450)$.

\begin {figure}  [htb]
\begin {center}
\vskip -16mm
\epsfig{file=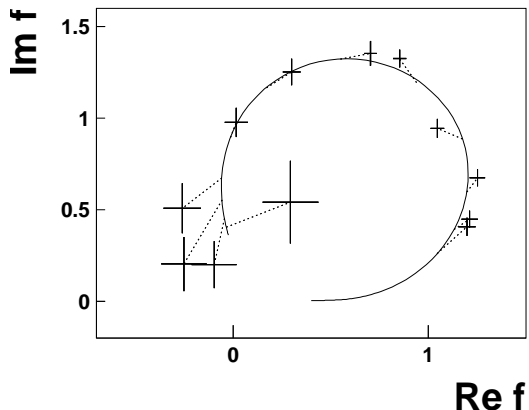,width=8.0cm}\
\vskip -6mm
\caption{The Argand loop of $a_0(1450)$ (full curve) compared with
magnitude and phase for individual bins 30 MeV wide, centred at
1.33 to 1.66 GeV.
Dotted lines show how individual bins move from the analytic
formula.}
\end {center}
\end {figure}

Finding the pole position of $a_0(1450)$ requires parametrisations of
$a_0\sigma$ and $\rho \omega$ phase space and the dispersive term in
the Breit-Wigner denominator.
This has been done with three alternative parametrisations for each of
the three terms.
Formulae are chosen with good convergence properties for complex $s$,
i.e. powers of $s$ confined to the denominators of the formulae.
All combinations of the formulae agree within $\sim 3$ MeV for both
real and imaginary parts of the pole, showing that systematic errors
for the extrapolation are well under control.
The pole position is $1432 - i98$ MeV;
the main systematic errors arise from the mean mass and width of
the peak.

Fig. 9 shows as dashed curves the intensities  of $a_0(1450) \to
a_0(980)\sigma$ and $\omega \rho$ as they appear in production
from $\bar pp$.
The $a_0(980)\sigma$ decay peaks at 1458 MeV and the $\omega \rho$
decay peaks at 1476 MeV.
Curves are normalised to 1 at their peaks.
Full curves show the peaks for an isolated resonance without the
limitation of $\bar pp$ phase space for
production.
The $a_0(980)\sigma$ peak is then at 1467 MeV and the
$\omega \rho$ peak at 1485 MeV.

\begin {figure}  [htb]
\begin {center}
\vskip -14mm
\epsfig{file=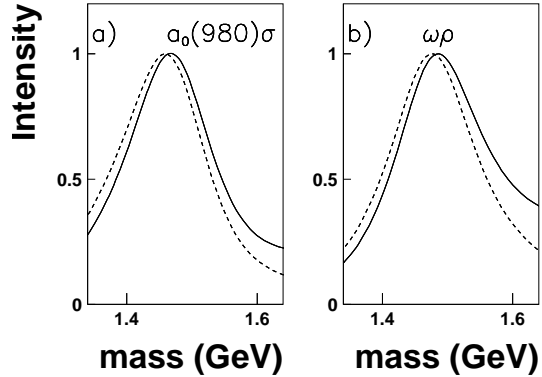,width=8.0cm}\
\vskip -4mm
\caption{Line-shapes of $a_0(1450)$ as it appears in decays
to $a_0(980)\sigma$ and $\omega \rho$, all normalised to 1
at the peaks. Dashed curves include the phase space for production in
$\bar pp$ annihilation; full curves are for an isolated resonance
without the limitation of the production process.}
\end {center}
\end {figure}

\subsection {Fitted parameters}
Table 2 collects results for $a_0(1450)$ from the final fit.
At this point, it is necessary to present a more convenient formula
for $a_0(1450)$ than Eq. (9), and the rationale behind it.

The basic points spring from the fact that there is a pole at
$1432 -i98$ MeV.
If one knew in advance how parameters vary between the pole and
the physical region, it would be best to write the formula directly
in terms of the pole and its residues, which express its coupling to
every channel.
That is not the case, so the closest approach is to write the formula
in terms of the nearby peak mass, $m_p = 1448$ MeV and widths to each
channel at this mass, together with their $s$-dependence:
\begin {eqnarray}
f(1450)&=& 1/[M^2 - s - i[\Pi (s) - \pi (M^2)] - im_p\sum _j
\Gamma _j (s)], \\
m_p\Gamma _j(s) &=& g^2_j \rho _j(s).
\end {eqnarray}
This form is close to that for a Breit-Wigner resonance of constant
width and is closely related to observed branching ratios between
channels.

The branching fractions for an isolated resonance are given by integrals
of the form
$$ \int \frac {g^2_j \rho _j(s)\, ds}{|D(s)|^2}. $$
If $\rho_j$, hence $\Gamma _j(s)$, were to vary linearly with $s$, the
variation of branching fractions would cancel between upper and lower
halves of the peak.
It turns out that this cancellation works fairly well.
This form of parametrisation gives a clear insight into the way the
fit responds to each parameter.

However, one important point emerges.
The sum of the widths at the peak comes to 345 MeV, considerably
larger than the observed full width of the peak, 192 MeV.
The reasons for this are straightforward.
On the lower side of the peak, $\Gamma (a_0(980)\sigma)$ and
$\Gamma (\omega \rho )$ are small, and the amplitude falls rapidly
because the remaining width to $\eta \pi$, $KK$ and $\eta '\pi$ is
small.
On the upper side of the peak, $\Gamma (a_0(980)\sigma )$ and
$\Gamma (\omega \rho )$ rapidly become large and dominate the
denominator, cutting off the line-shape $1/|D(s)|^2$ quickly.
The line-shape of $a_0(980)$ in Fig. 6(b) serves as a second
example.
The upper part of the peak is attenuated rapidly by
$\Gamma _{KK}(s)$.
The lower part is not far from a Breit-Wigner
resonance of constant width.

\begin{table}[htb]
\begin {center}
\begin{tabular}{l r}
\hline
Peak mass  & $1448 \pm 13 \pm 25$ \\
M(Breit-Wigner)  & $1536 \pm 20 \pm 30$ \\
Mean mass   & $1424 \pm 13 \pm 25$ \\
Full width at half maximum  & $192  \pm 9 \pm 9$ \\
Pole position  & $1432 \pm 13 \pm 25 - i(98 \pm 5 \pm 5)$ \\
$\Gamma (\eta \pi )$  & $23.7 \pm 0.5 \pm 2.0$ \\
$\Gamma (KK)$  (fixed from Eq. 11) & $17.7 \pm 0.3 \pm 2.0$ \\
$\Gamma (\eta '\pi )$  (fixed from Eq. 10) & $11.4 \pm 0.2 \pm 1.5$\\
$\Gamma (\omega \rho )$    & $219 \pm 18 \pm 24$ \\
$\Gamma (a_0(980)\sigma)$  & $73 \pm 5 \pm 20$ \\
In $\bar pp \to a_0(1450) \pi$:  \\
BR($a_0(980)\sigma)/BR(\eta \pi)$ & $2.3 \pm 0.2 \pm 0.6$ \\
BR($\omega \rho)/BR(\eta \pi)$ & $7.6 \pm 0.6 \pm 1.2$ \\
Branching fraction in $\eta \pi ^0\pi ^0$: \\
(a) with interferences  & ($5.44 \pm 0.15 \pm 0.88)\%$ \\
(b) without             & ($3.48 \pm 0.13 \pm 0.58$)\% \\
Branching fraction in $\omega \pi ^+\pi ^- \pi ^+\pi ^-$ &  \\
(a) with interferences  & ($3.49 \pm 0.14 \pm 0.30)\%$ \\
(b) without             & ($4.86 \pm 0.19 \pm 0.42)\%$ \\\hline
\end{tabular}
\caption{Results for $a_0(1450)$ in units of MeV.
The first errors are statistical and the second systematic.}
\end{center}
\end{table}

The branching ratio $\Gamma (a_0(980)\sigma )/\Gamma (\eta \pi)$
is obtained from the combined fit.
It is consistent with the magnitude of the $a_0(1450)$ signal fitted
to $\eta \pi ^+\pi ^-\pi ^+\pi ^-$ data in Ref. \cite {Nana}
and with the upper limit of 4.3 established there for this ratio.
The Table uses branching ratios to $\omega \rho$ from
the combined fit to $\omega \pi ^+\pi ^- \pi ^0$ data, discussed in
detail in subsection 3.9.
The Table quotes in lines 12 and 13 the ratio of branching fractions
$BR(\omega \rho )/BR(\eta \pi)$ and $BR(a_0(980)\sigma)/BR(\eta
\pi)$ as they appear in $\bar pp$ annihilation.
These values  are better defined than those for an isolated resonance
because of uncertainty about its high mass tail, see Fig. 9.
Note that $BR(\omega \rho )/BR(\eta \pi)$
is 7.6 in Table 2, rather smaller than the ratio $\Gamma (\omega \rho
)/\Gamma (\eta \pi ) = 9.2$ at the peak of $a_0(1450)$.
This is because the $\omega \rho$ signal in $\bar pp \to a_0(1450)\pi$
is inhibited at high mass by the available phase space, as illustrated
in Fig. 9(b) by the dashed curve.
For an isolated resonance, the $\omega \rho$ branching fraction is
close to 9.2, but with  an unknown error depending on form factors.

In Table 2, the first errors are statistical and the second systematic.
Strong contributions to systematic errors arise from uncertainties in
branching fractions for $\bar pp \to \eta \pi ^0\pi ^0$ $(\pm 5.0\%$)
and $\omega \pi ^+\pi ^-\pi ^0$ $(\pm 7.9\%)$.
However, the largest error arises from the fact that interferences
within one set of data lead to a branching fraction
$\propto |\sum _i \Lambda _i f_i|^2$, rather than
$\sum _i |\Lambda _i f_i|^2$.
Here, the sum is over resonances, $\Lambda _i$ are coupling constants
and $f_i$ are amplitudes for each resonance.
The second of these quantities is derived from $\Lambda$ parameters
fitted to the first, as explained above in subsection 3.5.

There is a potentially large error from the interference between the
two components making up the $\pi \pi$ S-wave: the $\sigma$ pole term
and the elastic component.
Fortunately, the $\eta \pi ^0 \pi ^0$ data determine both relative
magnitudes and phases of these two contributions quite well.
However, it is necessary to add a systematic error to cover the
change to the fit if a further term is added to the parametrisation
of the $\pi \pi $ S-wave.
Here, it is chosen to be the elastic amplitude multiplied by $s$.
There is a further small contribution to systematic errors from
perturbations when small components are dropped from the fits, e.g.
the weak $\pi _1(1400)$ contributions in both $^1S_0$ and $^3P_1$
annihilation.
Finally, in view of the scatter of the last 4 points of Fig. 8 above
1540 MeV, a systematic error is included from changes in the fit if
$a_0(1450)$ is fitted to $\eta \pi ^0 \pi ^0$ data only up to 1540 MeV.

Table 2 includes systematic errors in fitting
$\omega \pi ^+\pi ^-\pi ^0$ data.
The evaluation of systematic errors for these data follows the same
procedure as for $\eta \pi ^0 \pi ^0$.
The final systematic errors are added in quadrature.
It is not correct to add them linearly, as is
sometimes done. The derivation of the Gaussian error distribution
depends on the convolution of many box-shaped distributions.

\subsection {A disagreement with Obelix}
The Obelix group has published two claims to observe an $a_0$ decaying
to $\eta \pi$ in the mass range 1290-1313 MeV \cite {Bertin} \cite
{Bargiotti}.
Such a resonance should be very conspicuous in Crystal Barrel data
through distinctive interference with $a_2(1320)$.
The fits reported here have been repeated (i) using an $a_0$ in this mass
range without $a_0(1450)$ and (ii) together with $a_0(1450)$.
When $a_0(1450)$ is removed from the fit, $\chi ^2$ (scaled to allow
for the mean $\chi^2 $ of 2.9 per bin) is worse by 528 for a reduction
of six fitting parameters.
This is an 18 standard deviation signal.
If its mass and width are moved down to the mass range 1200-1340 MeV
with a width $\le  120$ MeV, there is no optimum.
Instead the fit moves in a few iterations  towards parameters of
$a_0(1450)$, whatever line-shape is used for $a_0(1450)$.
If an extra $a_0$ is added in the mass range 1280-1340
MeV, there is only a small improvement in $\chi ^2$ and again no
optimum for parameters in the range claimed by Obelix.
The narrow width they claim $\sim 80$ MeV is similar to that of
$a_2(1320)$.
It appears likely that their signal was confused with P-state
annihilation to $a_2(1320)$.
The P-state annihilation is precisely identified in present work from
data in hydrogen gas.

\subsection {The fit to $\omega \pi ^+\pi ^-\pi ^0$ data}
There are 35,280 reconstructed events for these data with
$8.4\%$ experimental background, arising in the selection of the
narrow $\omega$.
The earlier analysis of these data is reported in detail in
Ref. \cite {rwpi}.
Dispersive effects were included fully and the new fit changes rather
little.
Table 3 lists the components in the fit and their significance
levels, measured by changes in log likelihood when each component is
removed from the fit and all others are re-optimised.
Values of $\chi ^2$ are twice those for log likelihood for the
large statistics available here.

\vskip -2mm
\begin{table}[htp]
\begin{center}
\begin{tabular}{clcr}
\hline
Initial states & Channel & Intensity (\%) & $\Delta $Ln L \\\hline
$^1S_0$ & $a_0(1450)\pi$,                       & 3.5 & 90 \\
        & $b_1(1235)\rho, ~b_1 \to \omega \pi$  &13.2 & 361 \\

        & $\pi _1(1600)\pi ,~\pi _1 \to [b_1\pi ]_{L=0}$& 6.6 & 71\\
        & $a_2(1320)\pi$                               & 2.0 & 18 \\
        & $a_2(1660)\pi $                              & 2.4 & 36 \\
\hline
$^3S_1$ & All $\pi _0 \pi$                    & 16.9 & 505\\
        & $a_1(1260)\pi $                     &  0.8 & 48 \\
        & $a_1(1260)\omega, ~a_1 \to \rho \pi$& 23.9 & 377 \\
        & $a_1(1640)\pi$                      & 5.1  & 271 \\
        & $a_1(1640)\pi, a_1 \to b_1\pi$      & 1.7  & 45 \\ &
        $\pi _1(1600)\pi,~\pi _1 \to [b_1\pi ]_{L=0}$ & 2.5 & 50 \\
    & $\omega (1420)\pi ,~\omega (1420) \to \omega \sigma $ & 1.2 & 8\\
    & $\omega (1420)\pi ,~\omega (1420) \to b_1\pi  $ & 3.3 &   5 \\
    & $b_1(1235)\sigma, ~b_1 \to \omega \pi    $      & 1.6 &  57 \\
    & $\rho (1450)\sigma,~\rho \to b_1\pi$            & 0.6 &  94 \\
\hline
$^3P_0$ & $\pi_0 \pi     $                       & 9.5 & 83  \\
\hline
$^3P_1,~^3P_2$ & $a_2(1320)   $                       & 2.5 & 38    \\
               & $a_2(1660)   $                       & 5.7 & 45    \\
\hline
\end {tabular}
\caption {Percentage contributions of each channel after the
background subtraction.
Decays are to $\omega \rho$ unless stated otherwise.
The final column shows changes in log likelihood when each channel is
removed from the fit and remaining contributions are re-optimised.}
\end{center}
\end{table}
In the earlier work, there was a very marginal signal due to
$a_1(1260) \to \omega \rho$, which improved log likelihood by 24.
In the latest work, it improves log likelihood by only 6 and is
omitted from the fit.
Likewise the earlier work included a rather marginal signal for
$\pi(1600) \to (b_1(1235)\pi )_{L=2}.$
This contribution is now small and is set to zero.

A further detail is that there are data for
$\bar pp \to \omega \pi ^0\pi ^0\pi ^0$ \cite {Rod}.
The branching ratio for this channel is very small.
These data constrain the magnitudes of the last four entries to
Table 3 for $^3S_1$.
Their phases are fitted freely.
Two of them have only very small effects in the present fit.

Let us review the essential points of the analysis.
There are three charge combinations of $\pi \pi$.
As a result, individual resonances do not appear clearly in mass
projections.
It is necessary to rely on the amplitude analysis to locate magnitudes
and phases from what it finds in 4-body phase space.
That may appear questionable, but in practice works well.
The fits to mass projections were shown in Fig. 2 of the earlier paper.
The tiny changes in the new fit are hardly visible by eye and
therefore the figures will not be repeated here.

Secondly, angular distributions depend distinctively on spins.
Consider $a_0(1450)$ as an example.
The spin of the $\omega$ lies along the normal to its decay plane.
The spin of the $\rho$ is given by the vector $(k_1 - k_2)$, where
$k_{1,2}$ are momenta of the pions from its decay.
After Lorentz transformations to the $\omega \rho$ rest frame, the
matrix element is given by the scalar product of these two vectors.
This is highly distinctive.
An elementary check on formulae is that all amplitudes are orthogonal.
One can test how well quantum numbers are recognised by putting
deliberate errors into formulae.
Generally the result is that the amplitudes drop to small values.

A third point is that the programme prints a matrix giving intensities
of all components together with real and imaginary parts of all
interferences.
This identifies the important interferences.
It is then easy to test the reliability of these interferences by
plotting log likelihood against relative phases.

The magnitude of the $a_0(1450)$ signal has decreased slightly from the
earlier publication, but its significance level has improved.
In the earlier work, log likelihood changed by 56 when $a_0(1450)$ was
omitted from the fit.
Now it changes by 90.
It is produced from the $^1S_0$ initial state where it interferes with
a  large and well identified $b_1(1235)\rho$ signal.

There is a large signal in $^3S_1$ annihilation from
$J^{PC} = 0^{-+}$.
It peaks at 1480 MeV, quite close to $a_0(1450)$.
One might worry that there will be cross-talk with $a_0(1450)$,
despite the fact that the $a_0(1450)$ is produced from the $^1S_0$
initial state while $0^{-+}$ is produced from $^3S_1$.
There is no such problem.
The $0^{-+}$ component may be removed completely from the fit
without any significant effect on the fitted $a_0(1450)$ signal.
In fact, the $a_0(1450)$  is insensitive to changes in all $^3S_1$,
and P-state amplitudes.

\begin {figure}  [htb]
\begin {center}
\vskip -12mm
\epsfig{file=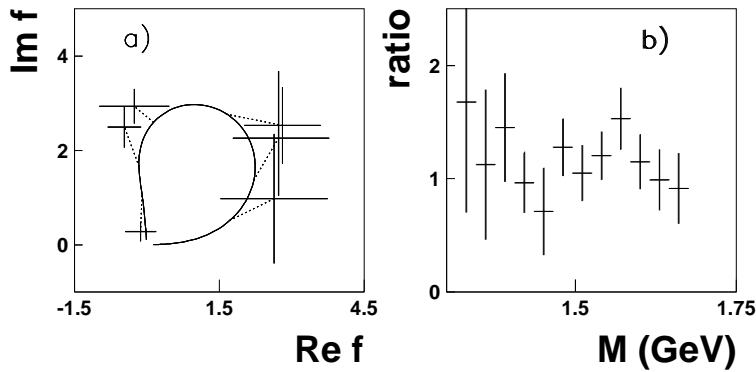,width=11.0cm}\
\vskip -3mm
\caption{(a) The Argand loop of $a_0(1450)$ in $\omega \rho \pi$ (full
curve) compared with magnitude and phase for individual bins 60 MeV
wide, centred at 1.345 to 1.645 GeV.
Dotted lines show how individual bins move from the analytic formula.
(b) The ratio of the $a_0(1450)$ amplitude in bins 30 MeV wide to that
of the overall fit.}
\end {center}
\end {figure}
A similar comparison has been made in Fig. 10(a) to that shown in
Fig. 8.
Real and imaginary parts of the $a_0(1450) \to \omega \rho$ signal are
fitted freely in bins 60 MeV wide from 1.345 to 1.645 GeV.
In Fig. 10(a), the amplitude is $k'\sqrt {\rho _{\omega \rho}}/D(s)$;
i.e. it allows for the phase space of the $\omega \rho$ final state
and the phase space in the production reaction
$\bar pp \to a_0(1450) \pi$, proportional to the momentum $k'$ of the
$a_0(1450)$ in the $\bar pp$ centre of mass.
There is no doubt that the data conform with a resonant circle, though
errors are sizable.
There appears to be some tendency for the data to require a larger
amplitude than the overall fit.
An alternative test is made by fixing the phase of the amplitude to
that of the overall fit, but allowing the magnitude of the fitted
signal to fit freely in 30 MeV wide bins.
Fixing the phase stabilises the fitted amplitude considerably.
Results are shown in Fig. 10(b). There is now a reduced tendency for
the fitted amplitude to be above the overall fit.
The mean discrepancy is $(13.7 \pm 8.5)\%$.
The systematic difference arises because the final fit is constrained to
fit the line-width of the $a_0(1450)$.
If the $a_0(1450) \to \omega \rho$ signal is increased, it makes the
line-width smaller; this was the problem with the first publication,
Ref. \cite {rwpi}.
The final fit is a compromise between fitting the line-shape and the
magnitude of the $a_0(1450) \to \omega \rho$ signal.

In assessing the errors for $\Gamma (\omega \rho )/\Gamma (\eta \pi )$
in Table 2, the statistical error is taken from the
$8.5\%$ statistical error in the discrepancy of Fig. 10(b).
This is quite close to the error derived from log likelihood in
Table 3.
The systematic error is derived from changes in
$\Gamma (\omega \rho )/\Gamma (\eta \pi )$ as the mass and width of
$a_0(1450)$ are varied over the range of systematic errors in Table 2.

One new point does emerge from a better understanding of dispersive
effects.
This concerns the large $J^P = 0^{-+}$ component.
In the earlier work, attempts were made to fit it with $\pi (1300)$
and a radial recurrence in the mass range 1600-1700 MeV.
However, the required signal for $\pi (1300)$ was unreasonably large
and would have required it to decay dominantly to $\omega \rho$.
Furthermore, the data still required a definite peak in the vicinity of
1500 MeV.
A radial recurrenece so close to $\pi (1300)$ would be surprising.

The present work reveals a more sensible way of fitting the $0^-$
signal.
The optimum fit is obtained with a broad resonance at 1540 MeV with
$\Gamma = 590$ MeV, plus a radial excitation of $\pi (1300)$ at
$1732 \pm 32$ MeV with $\Gamma = 252 \pm 30$ MeV.
The broad resonance is close to being a simple cusp at the
$\omega \rho$ threshold.
This solution is shown in Fig. 11.

\begin {figure}  [htb]
\begin {center}
\vskip -16mm
\epsfig{file=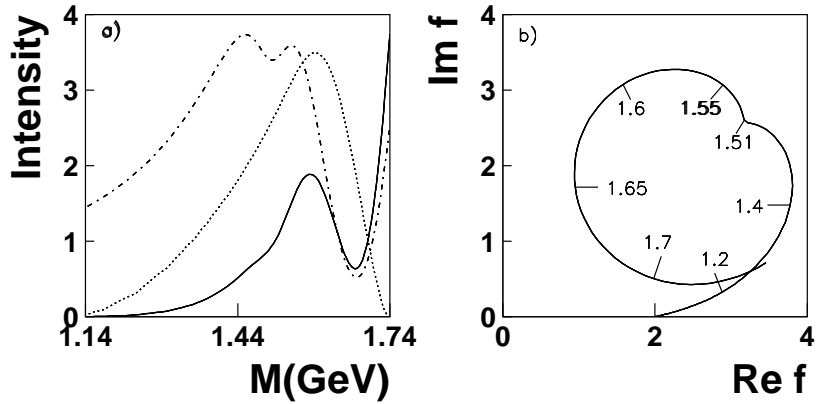,width=12.0cm}\
\vskip -3mm
\caption {(a) The line-shape of the $J^{PC} = 0^{-+}$ signal as it
would appear in $\omega \rho$ without the limitation of production in
$\bar pp \to \omega \rho \pi$ but (i) including
$\omega \rho$ phase space (full curve) and (ii) with this phase space
factored out (chain curve).
The dotted curve shows the phase space for $\omega \rho $ including the
limitation due to production in $\bar pp \to \omega \rho \pi$.
(b) The Argand diagram; masses are shown in GeV. Units are arbitrary
for display purposes.}
\end {center}
\end {figure}

Fig. 11(a) shows as the full curve the intensity of the fitted $\omega
\rho$ signal including $\omega \rho$ phase space, for a resonance
`in free space', i.e.  without the limitation imposed by production in
$\bar pp \to \omega \rho \pi$.
The phase space for $\bar pp \to \omega \rho \pi$ including this
limitation is shown by the dotted curve.
What is actually observed in $\omega \pi ^+\pi ^-\pi ^0$ data is then
given by the product of the dotted curve and the full curve.
The chain curve shows the result of dividing the full curve by
$\omega \rho $ phase space.
This result shows the line-shape arising from $|\Lambda
_A/D_A(s) + \Lambda _B/D_B(s)|^2$, where $A$ and $B$ refer to the two
components and $\Lambda$ are coupling constants.
This is what one would see for an isolated resonance if there were no
phase space factor in the numerator.
The double peak near 1.45 GeV comes from the cusp +
interference.
The peak at high mass comes from the radial recurrence at 1732 MeV.

Minor variants on this solution are possible because the $\omega \rho$
amplitude below 1450 MeV is small and poorly determined, particularly
its phase.
One should therefore not place any reliance on the threshold behaviour
of the $\omega \rho$ amplitude below 1450 MeV.
It is quite possible that the double peak at 1430 and 1480 is an
artefact.
However, there are two essential features which are unavoidable.
The first is a peak at $\sim 1550$ MeV in $\omega \rho$.
This arises from the cusp at the $\omega \rho$ threshold.
The second well determined feature is the phase advance $>180^\circ$
from 1510 to 1740 MeV.
This arises largely from the radial recurrence at 1732 MeV.

The present data are limited by the fact that
production of $0^-$ from $^3S_1$ is suppressed at the highest
masses by the $L=1$ centrifugal barrier for production.
A quite significant signal is however observed also in P-state
production with $L=0$.
This signal gives a reasonable determination of the mass and width
of the upper resonance at 1732 MeV.

There is some chance that this radial excitation
corresponds to $\pi (1800)$.
However, using PDG parameters for $\pi (1800)$, the fit is 2.8
standard deviations worse than with a free fit.
The mass and width observed in the present fit correspond closely to
those observed by Amelin et al. in $\pi ^- A \to \omega \pi^-\pi ^0 A^*$
\cite {Amelin}.
The ideogram shown by the PDG for the mass of $\pi (1800)$
has a double-humped structure.
There is the possibility of a $0^-$ hybrid
in this mass range to accompany the $\pi_1(1600)$.
So there is room for a conventional radial excitation around
1730 MeV and a hybrid at higher mass.
The $\pi (1800)$ has decay modes suggestive of a hybrid.
A fit using a mass and width from the higher lobe of the PDG's ideogram
is worse than the free fit by 4.0 standard deviations.
Further exploration of $0^-$ signals is needed in this mass range to
resolve the current uncertainties.

\begin {figure}  [htb]
\begin {center}
\vskip -16mm
\epsfig{file=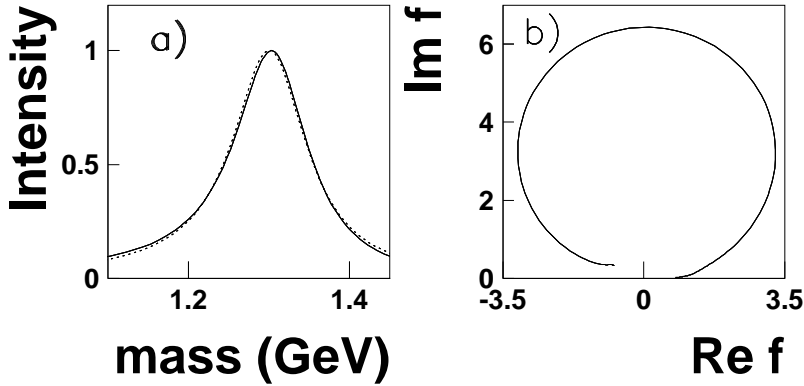,width=12.0cm}\
\vskip -3mm
\caption {(a) The line-shape of $a_2(1320)$: full curve including
$s$-dependence of the width and dispersive effects in full, the
dotted curve showing a Breit-Wigner amplitude of constant width.
(b) The Argand diagram. }
\end {center}
\end {figure}

A final detail concerns $a_2(1320)$.
Fig. 12 illustrates the small effect on $a_2(1320)$ of the
$s$-dependence of the width and dispersive effects.
The full $s$-dependence of decays to $\rho \pi$, $\eta \pi$, $KK$ and
$\omega \rho$ is included.
The full curve on Fig. 12(a) shows the fitted line-width and the dotted
curve the line-shape of a Breit-Wigner amplitude of constant width
agreeing at half-height with the full curve.
There is little difference between them, showing that the slow
$s$-dependence of the dominant $\rho \pi$ channel has little effect.
Fig. 12(b) shows that the Argand diagram follows a circle closely.

\section {Results for $a_0(980)$}
In the  1994 work \cite {1450B}, the $a_0(980)$ was fitted with
a Flatt\' e formula with $M = 999 \pm 5$ MeV, $g^2(\eta \pi ) =
221 \pm 20$ MeV and $g^2(KK)/g^2(\eta \pi ) = 1.16 \pm 0.18$.
The $\eta' \pi$ channel was not included.
These parameters now change beyond their errors because (a)
the large dispersive cusp at the $KK$ threshold is included, (b)
Adler zeros are included in both $\eta \pi$ and
$\eta '\pi$ channels.
These are at $s = s_A = m^2_\eta - 0.5 m^2_\pi = 0.2905$ GeV$^2$ for
the $\eta \pi$ channel and at $s = s_{A'} = m^2_{\eta '} - 0.5 m^2_\pi
= 0.9078$ GeV$^2$ for the $\eta '\pi$ channel.
Note that the latter is quite close to the resonance mass.

The formulae used for coupling to $\eta \pi$ and $\eta '\pi$ are
\begin {eqnarray}
g^2_{\eta \pi }(s) &=& g^2_{\eta \pi }(4m^2_K)\frac {s - s_A}{4m^2_K -
s_A} \exp (-2\alpha k^2_\eta ), \\
g^2_{\eta ' \pi }(s) &=& g^2_{\eta ' \pi } (m_{\eta '} + m_\pi )^2
\frac {s - s_{A'}}{(m_{\eta '} + m_\pi)^2 - s_{A'}}
\exp (-2\alpha k_{\eta '}^2),
\end {eqnarray}
where $k_{\eta '}$ is the momentum in the $\eta '\pi$ rest frame.
The value of $g^2_{\eta '\pi}$ is normalised at the $\eta '\pi$
threshold.

Below the thresholds for these processes, the Flatt\' e formula
has the sub-threshold analytic continuation
\begin {equation}
g^2_{\eta ' \pi}\rho _{\eta '\pi} \to i g^2_{\eta '\pi} 2|k^2_{\eta
'}|/s.
\end {equation}
However, $g^2_{\eta '\pi}$ becomes real again for a mass $< 0.83$ GeV,
due to the opening of the $u$-channel.
It makes little sense to allow for this without including the dynamics
of the $u$-channel process.
Therefore $g^2_{\eta '\pi}$ is set to zero below $s = s_{A'}$.
The effect of the $\eta '$ channel on $\chi^2$ of the fit is quite
small, except above the $\eta '\pi$ threshold.
Its coupling constant is fixed to that of the $\eta \pi$ channel by Eq.
(10).
Note that the assumption is made in Eq. (16) that the form factor
does not affect the ratio $g^2_{\eta '\pi}/g^2_{\eta \pi }$ between
these two thresholds.
Because of the small effect of the $\eta' \pi$ channel on present fits,
this assumption has little effect on $\chi^2$.
However, it could matter if and when data become available directly
on the $\eta '\pi$ channel.
A detail is that the opening of the $\eta '\pi$ channel is visible
on the Argand diagram of Fig. 7, just below 1.1 GeV.

For the $KK$ channel, the Adler zero is far away at $s = 0.5 m^2_K$
and experience with the $\sigma$ amplitude is that the factor
$(s - s_A)/(4m^2_K - s)$ needs to be multiplied by an exponential form
factor $\exp (-\alpha s)$, which prevents the amplitude rising
indefinitely with $s$.
Below the $KK$ threshold, the $KK$ channel has only an indirect effect
on data in the $\eta \pi$ channel.
Tests have been made with a variety of form factors.
Within errors, the best fit is obtained with $g^2_{KK}$ = constant
below the $KK$ threshold and this simple prescription has been adopted.
Above the $KK$ threshold, the factor due to the Adler zero is
dropped and the form factor $\exp (-2 k^2_{KK})$ is used, with $k_{KK}$
the momentum in the $KK$ channel in GeV/c.

\subsection {Fits to $\eta \pi ^0 \pi ^0$ data}
The dispersive cusp locks the mass $M$ of the amplitude at or just
below the $KK$ threshold and provides considerable stability.
Values of $g^2_{\eta \pi}$ and the ratio $r_{KK} = g^2_{KK}/g^2_{\eta
\pi }$ are only slightly correlated in fitting $\eta \pi \pi$ data;
there is a weak tendency ($5\%)$ for them to go up and down together.
The data give well defined values for parameters of $a_0(980)$:
\begin {eqnarray}
M &=& 0.9874 \pm 0.0010(stat) \pm 0.0030(syst)~ GeV, \\
g^2_{\eta \pi}&=& 0.164 \pm 0.007 \pm 0.010~GeV^2, \\
r_{KK} = \frac {g^2_{KK}}{g^2_{\eta \pi }} &=& 1.05 \pm 0.07 \pm 0.05.
\end {eqnarray}
These values have changed from earlier publications because of the
inclusion of the cusp in ${\rm Re}\, \Pi(s)$.
The `edge' observed in the $\eta \pi \pi$ data due to
$a_0(980)$ provides a good determination of the coupling to
$KK$.
In this respect, Crystal Barrel data have an advantage over Kloe data
(to be discussed further below).
The Kloe data however provide an excellent view of the $a_0(980)$
line-shape below the $KK$ threshold.
A simple program evaluating the formulae for $a_0(980)$  and the cusp in
$\rm {Re}\, \Pi (s)$ is available from the author.

The systematic error on the mass $M$ arises from uncertainty in the
mass calibration of the Crystal Barrel detector.
Systematic errors for $g^2_{\eta \pi}$ and $r_{KK}$ arise as described
above for the entire fit to $\eta \pi ^0\pi ^0$ data.
The fitted mass of $a_0(980)$ is close to the lowest $KK$ threshold,
just as the mass of $X(3872)$ is close to the threshold of the lowest
charge combination in $\bar D D^*$.

An important detail is that the $a_0(980)$ could be produced either
by its pole term or via the elastic $\eta \pi$ amplitude.
Both have been tried, and the fit strongly prefers production
via the elastic scattering amplitude, i.e. with the Adler zero in
the numerator of the production amplitude.

\begin {figure}  [htb]
\begin {center}
\vskip -16mm
\epsfig{file=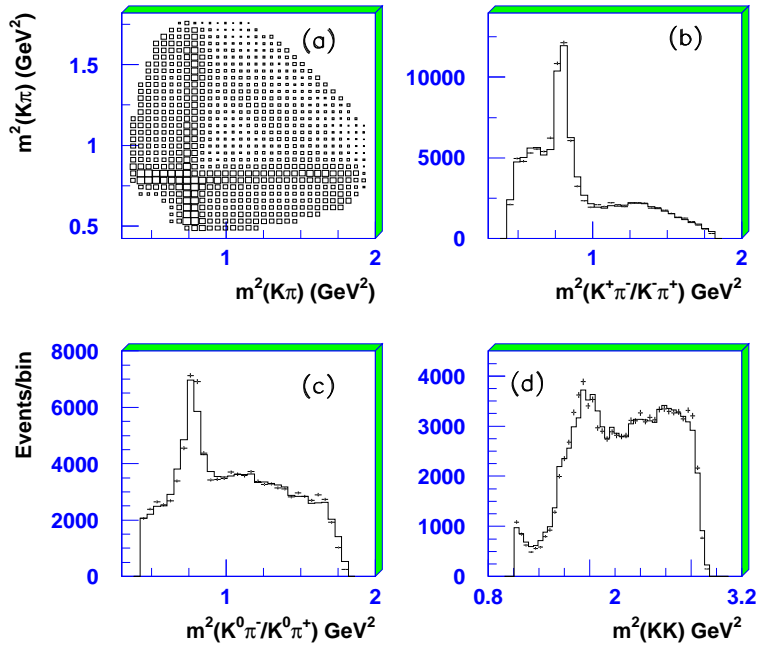,width=10.5cm}\
\vskip -3mm
\caption {Data and fits for $\bar pp \to K^0_L K^\pm \pi^\mp$ at
rest. (a) Dalitz plot; mass projections for (b) $K^+\pi ^-$ and
$K^-\pi ^+$, (c) $K^0_L\pi ^-$ and $K^0_L\pi ^+$, (d) $K^0_L K^\pm$.
Points with errors show data and histograms show the fit.}
\end {center}
\end {figure}

\section {Data for $\bar pp \to K^0_L K^\pm \pi ^\mp$}
The Dalitz plot for these data in liquid hydrogen is shown in
Fig. 13.
There are prominent vertical and horizontal bands due to $K^*(890)$.
A detail is that it is necessary to fine-tune the masses and widths of
the separate charge states for $K^*(890)$.
There are also diagonal bands due to $a_2(1320)$ and $a_0(980)$.
The $a_0(1450)$ lies in the lower left corner of the plot, near the
crossing $K^*(890)$ bands.
As for $\bar pp \to \eta \pi ^0\pi ^0$, it is necessary to remove
some edge bins.
There is also a background from $\bar pp \to \pi ^+\pi ^-\pi ^0$
described in the Crystal Barrel publication.
It peaks in edge  bins close to the left-hand corner of
the Dalitz plot.
It is necessary to remove 21  bins, leaving 741.

One would hope to determine the ratio $g^2_{KK}/g^2_{\eta \pi}$
for $a_0(980)$ from relative contributions in $\eta \pi \pi$ and
$K\bar K\pi$ data.
Unfortunately, when one tries to do this, a serious difficulty appears.
It arises from the question of how to parametrise the $K\pi$ S-wave
amplitude in the $K\bar K\pi$ data.
This leads to uncertainties in interferences between
the $K\pi$ S-wave and  diagonal bands due to $a_2(1320)$,
$a_0(980)$ and $a_0(1450)$.
Uncertainties in these interferences then lead to uncertanties in the
magnitudes of the $a_2$ and $a_0$ signals.
To grasp these points, it is necessary to review current understanding
of the $K\pi$ S-wave, which has advanced a long way since the earlier
analysis of the Crystal Barrel data \cite {Spanier}.

\begin {figure}  [htb]
\begin {center}
\vskip -16mm
\epsfig{file=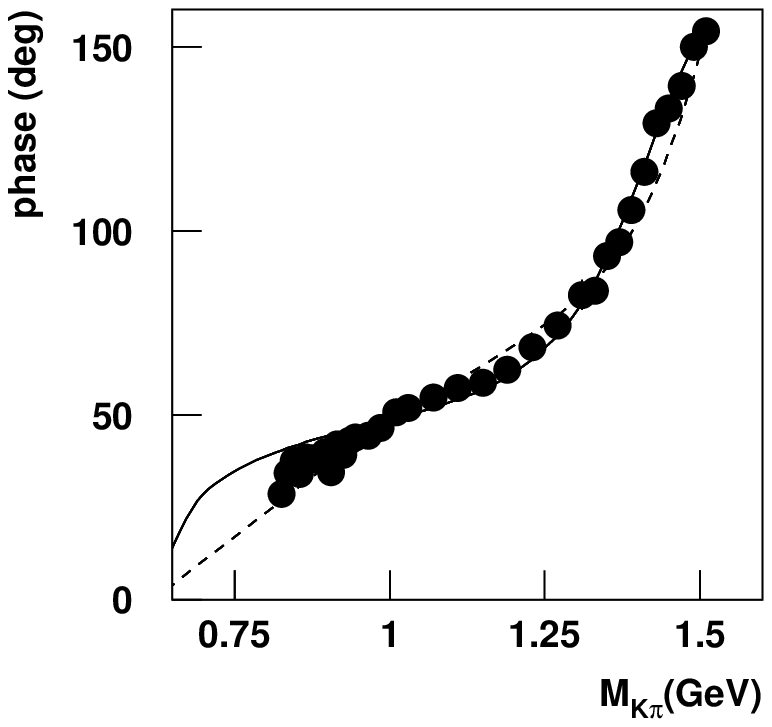,width=8.0cm}\
\vskip -6mm
\caption {LASS data for $K\pi$ elastic scattering, compared with
(i) the Crystal Barrel fit (full curve), (ii) a parametrisation
including the Adler zero in $K\pi$ (dashed curve).}
\end {center}
\end {figure}

In that early analysis, it was assumed that the $K\pi$ S-wave
amplitude in production data is identical to that in $K\pi$ elastic
scattering.
Experience with $\pi \pi$ data now makes that appear unlikely \cite
{Extended}.
The Crystal Barrel paper gives an explicit parametrisation for the
$K\pi$ S-wave.
Fig. 14 shows its  phase as a function of mass, compared to LASS data
\cite {LASS}.
In  those days, the LASS data were fitted with an effective range
expression.
Since then, it has been recognised that Chiral Symmetry breaking
produces an Adler zero in the $K\pi$ S-wave below threshold at
$s = m^2_K - 0.5m^2_\pi$.
The dashed curve shows the amplitude fitted to the LASS data
including this Adler zero \cite {Bugg}.
The $\kappa$ pole observed in BES2 \cite {kappa} and E791
\cite {E791} data was fitted simultaneously and is therefore well
constrained.
There is a discrepancy for $K\pi$ masses near 1400 MeV.
It is now known that this discrepancy can be removed by a full
treatment of the cusp at the $K\eta '$ threshold.
For present purposes it is irrelevant since $KK\pi$ phase space ends
at 1381 MeV.

The discrepancy near the $K\pi$ threshold between the dashed and full
curves on Fig. 14 indicates a contribution in Crystal Barrel data from
the $\kappa$ pole, which peaks near threshold with a half-width of
$\sim 350$ MeV.
This is confirmed by a fit to the Crystal Barrel data using a
2-component fit to the $K\pi$ S-wave, as in Eq. (8).
It contains one component from $K\pi$ elastic scattering and a second
from the $\kappa$ pole.
Uncertainties are compounded by the fact that there are contributions
from $\kappa \pi$ for both $I=1$ and $I=0$ initial
states, making four $K\pi$ S-wave amplitudes in all.
These two isospins have opposite relatives signs for coupling
to $K\pi$ and are responsible for the difference in $K\pi$
distributions between Figs. 13(a) and (b).

There is in addition the possibility of a third component due to
a $K_0(1430)$ amplitude different in magnitude and phase in
production data and LASS data.
That is the case for the E791 data.
However, it turns out that adding this freedom only improves $\chi^2$
by a small amount, $\sim 20$ and leads to a very ill-defined fit.
This extra possible freedom is ignored here.

The earlier Crystal Barrel analysis recognised the need for a low mass
$K\pi$ enhancement and parametrised it in an {\it ad hoc} way which is
not consistent with Chiral Symmetry breaking and the Adler zero.
In order to get an acceptable fit, contributions were introduced
from $\rho (1450)$ and/or $\rho (1700)$, which can appear
in the left-hand corner of the Dalitz plot, in the same mass range
as the $\kappa$ pole and $a_0(1450)$.
It is now necessary to try to disentangle the complications of this
corner of the Dalitz plot.
Unfortunately, these complications lead to substantial errors in
branching ratios.

Results will be presented first without any contribution from
$\rho (1450)$ and $\rho (1700)$; then their possible contributions
will be discussed.
Either way, there is a large interference between the $\kappa $ pole
and the $K\pi$ elastic amplitude.
With this freedom, there is great flexibility in what can be fitted
to the $K\pi$ S-wave.
A free fit to $a_0(980)$ and $a_2(1320)$ magnitudes gives $\chi^2 =
637.3$ for 720 degrees of freedom.
If the $\kappa$ pole is omitted, $\chi^2 \to 1230$; this is
clearly unacceptable.
The free fit gives a ratio of intensities $a_0(980)/a_2(1320) = 0.0637$
compared with a prediction from fits to $\eta \pi \pi$ data,
Eqs. (18-20), giving a ratio $0.0369 \pm 0.0068$.
This might appear to be a large discrepancy.
However, if the fit is constrained to agree with the latter prediction,
$\chi^2$ increases by only 2.01 to 639.31, for 2 parameters determining
the complex coupling constant $\Lambda$ of $a_0(980)$.
There is clearly no disagreement with the prediction; there
is just a large flexibility in the amplitudes.

The same picture emerges for the ratio of intensities for $a_0(1450)/
a_2(1320)$.
A free fit to $a_0(1450)$ gives a ratio of intensities
$a_0(980)/a_2(1320) = 0.535$ compared with its predicted value 0.311.
However, again the ratio can be fixed to the prediction with a $\chi^2$
increase of only 2.01 for 2 less parameters for $\Lambda _{1450}$.

With the introduction of extra contributions from $\rho (1450)$ and/or
$\rho (1700)$ the situation is similar.
If both are introduced, the solution becomes very unstable with
excessively large contributions from both resonances and destructive
interference between them.
This destructive interference is a familiar symptom of over-fitting
the data, so it is necessary to use only one of them.
That procedure was adopted in the earlier Crystal Barrel analysis.
There are recent data from Babar \cite {Babar} on $e^+e^- \to
\gamma K^+K^-\pi ^0$ using Initial State Radiation.
They observe the same instability between $\rho (1450)$ and $\rho
(1700)$ contributions, but a dominant $\rho (1450)$ amplitude.
Following this lead, the present data are fitted including only
$\rho (1450)$.
The story which then emerges runs close to that described above.
For a fit where $a_0(1450)$ and $a_0(980)$ are constrained to
predictions from Eqs. (18-20), $\chi^2$ drops from 639.31
to 616.50 with the addition of $a_0(1450)$.
If $a_0(980)$ is set free, $\chi ^2$ improves by 3.09 to 613.41.
This cannot be regarded as a significant improvement.
If $a_0(1450)$ is set free, the improvement is only 1.74, again
insignificant.

So the conclusion is that data on $\bar pp \to K^0_LK^\pm \pi ^\mp$
are consistent within errors with (i) the parameters of $a_0(980)$
deduced from $\eta \pi \pi$ data, (ii) the SU(3) prediction for
$g^2_{KK}/g^2_{\eta \pi}$ of $a_0(1450)$.
However, they do not constrain those parameters well.
A close inspection of the $KK$ mass projection on Fig. 13(d) shows that
data favour a slightly narrower $a_0(980)$ than is fitted.
This could arise from either stronger coupling of $a_0(980)$ to $KK$
or a steeper form factor, i.e. a larger $\alpha$ parameter and larger
radius of interaction.

A final question is whether there is any significant $a_0(1450) \to KK$
signal at all.
This may be tested against the fit where $a_0(1450)$ and $a_0(980)$
are constrained to their predicted values.
To err on the pessimistic side, this test is made including $\rho
(1450) \to KK$.
Omitting $a_0(1450)$ from the fit, $\chi^2$ gets worse by 36.8 with 2
less parameters.
This is a change of 5.5 standard deviations, so there does appear
to be a signal due to $a_0(1450) \to KK$.
The situation is simply that its magnitude cannot be determined
with any precision from these data.

\subsection {Comparison with Kloe data on $\phi \to \gamma (\eta \pi)$}
There are important data from Kloe on this process giving direct
and precise information on the line-shape of $a_0(980)$.
The present status is that data have been published from the
first phase of this experiment, with $\eta $ decaying via both
$\gamma \gamma$ and $\pi ^+\pi ^-\pi ^0$ \cite {Kloe}.
Preliminary results for parameters of $a_0(980)$ have also been
presented from the second stage of this work \cite {Ambrosino}.
A comparison will be made with these preliminary results.

The Kloe data define accurately the $a_0(980)$ line-shape below the
$KK$ threshold.
In the process $\phi \to \gamma a_0(980)$, there is a dependence
of the cross-section on $k^3_\gamma$, where $k_\gamma$ is the
photon momentum.
This factor is the usual $k^3$ factor for an E1 transition.
It inflates the lower side of the $a_0(980)$ strongly, improving the
precision with which it can be measured.
However, one must beware that the line-shape may be affected by
form factors.

The Kloe group have fitted their data in two ways.
The first assumes the $KK$-loop model of Achasov and Ivanchenko
\cite {Ach}.
In this model, $\phi$ decays to $\gamma (KK)$, then the $KK$ pair
re-scatter via a final state interaction to $a_0(980)$, which
decays to $\eta \pi$.
With this model, the preliminary parameters for $a_0(980)$ are
\begin {eqnarray}
M &=& 983 \pm 1~MeV \\
g^2_{\eta \pi} &=& 0.156 \pm 0.011 ~GeV^2                \\
r_{KK} = \frac {g^2_{KK}}{g^2_{\eta \pi}} &=& 1.19 \pm 0.05.
\end {eqnarray}
These values are quite close to those emerging from Crystal Barrel
$\eta \pi \pi$ data, Eqs. (18-20).

The alternative fit made by Kloe ignores the constraint of the $KK$
loop model and arrives at a value of $g^2_{\eta \pi}$ seriously
different: $g^2_{\eta \pi} = 0.096 \pm 0.009$ GeV$^2$.
The agreement between the first set of results and Crystal Barrel data
clearly favours the $KK$-loop model.
The disagreement of the second set illustrates the sensitivity to the
precise equations used to fit Kloe data.

The published Kloe data were fitted in earlier work
using the $KK$ loop model \cite {Recon}.
That fit has now been repeated using the parameters for $a_0(980)$
reported here.
There is a significant  contribution to the data (up to $18\%$)
from $\phi \to \rho \pi$, $\rho \to \gamma \eta$.
From present publications, it is not clear how much of this
contribution is eliminated by experimental cuts.
Excellent fits can be obtained with both the Kloe parameters of
Eqs. (21-23) and with parameters fitting Crystal Barrel
data, Eqs. (18-20) by varying the magnitude and phase of the $\rho \pi$
combination.
The magnitude of the fitted $a_0(980)$ signal is
proportional to $g^2_{\eta \pi} g^2_{KK}$ and is constrained to
reproduce the latest branching ratios reported in \cite {Ambrosino}.
The fit is almost indistingushable from that shown in Fig. 3 of
\cite {Recon}.
When information from the full Dalitz plot becomes available, the
$\rho \pi$ amplitude can be determined accurately in both magnitude and
phase.

\subsection {Pole parameters of $a_0(980)$}
In order to find the pole position of $a_0(980)$ it is necessary
to parametrise the cusp at the $KK$ threshold in the real part of
the amplitude.
At the cusp, there is a discontinuity in slope, due
to the opening of the $KK$ channel.
If there is no form factor in the $KK$ channel, this cusp can
be calculated algebraically with two subtractions at the $KK$ threshold.
It is given in Ref.  \cite {sync}, Eqs. (8) and (9):
\begin {eqnarray}
\frac {\rm {Re} \, \Pi (s)}{g^2_{KK}} = j_{KK} &=&
\frac {\rho _{KK}}{\pi }
\rm {ln} \frac {1 - \rho_{KK}}{1 + \rho _{KK}}, \qquad s \ge 4m^2_{KK}
\\
&=& - \sqrt {\frac {4m^2_{KK} - s}{s}} - \frac {2v}{\pi }
\tan ^{-1} v,
\qquad s < 4m^2_{KK},
\end {eqnarray}
where $\rho_{KK} = 2k_{KK}/\sqrt {s}$ above the $KK$ threshold and
$v$ is the modulus of this quantity below threshold;
$k_{KK}$ is the momentum in the $KK$ rest frame, and is complex
below threshold.

With the form factor present, an empirical parametrisation is needed,
based on Eqs. (24) and (25).
An excellent fit may be obtained to $\Pi (s)$ from the $\eta \pi$
threshold to 1.6 GeV replacing these equations by
\begin {eqnarray}
\rm {Re} \, \Pi (s)   &=& f(s)\, j_{KK} + F_2 + F_3 s + F_8 s^2_p \\
f(s)   &=& \frac {F_1}{1 + F_4s_r + F_5s^2_r + F_7 s^3_r} \\
s_r     &=& s - 4m^2_{KK} \\
s_p     &=& s - F_6.
\end {eqnarray}
Here $f(s)$ modulates $j$ with a convergent power series about the
$KK$ threshold. [A similar expression is used for $a_0(1450)$].
The terms $F_2 + F_3 s$ allow for the double subtraction at the $KK$
threshold.
The term $F_8s^2_p$ is mostly concerned with fitting
$\rm {Re} \, \Pi(s)$ above $s = 1.5$ GeV$^2$.
Parameters are tabulated in Table 4.

\vskip -4mm
\begin {table} [htb]
\begin {center}
\begin {tabular}{cc}
\hline
$F_1$ &  0.722107 \\
$F_2$ &  0.052635 \\
$F_3$ & -0.230099 \\
$F_4$ & -0.638335 \\
$F_5$ &  1.804376 \\
$F_6$ &  2.165924 \\
$F_7$ &  1.402098 \\
$F_8$ &  0.172984 \\\hline
\end {tabular}
\caption {Parameters (in units of GeV) fitted to Eqs. (26)--(29).}
\end {center}
\end {table}

The physical region lies at $s + i\epsilon$ for all channels,
in the limit $\epsilon \to 0$.
The $\eta \pi$, $KK$ and $\eta '\pi$ cuts may be labelled by the signs
multiplying $i$ for each channel.
The pole closest to the physical region has signs $+-+$ and lies at $M
- i\Gamma /2 = 989.1 - i 40.1$ MeV.
It is reached from the physical region by going around the end of the
$KK$ cut and is usually called the second-sheet pole.
Tests show that this pole moves little with the $KK$ form factor.
The strong coupling to $KK$ locks the resonance to the $KK$ threshold
and the full width $\Gamma = 80.2$ MeV is determined by the fitted
value of $g^2_{\eta \pi }$.
If the sign of $i$ for the $\eta '\pi$ channel is reversed, the pole
moves to $997.4 - i46.3$ MeV.
The change due to the $\eta '\pi$ channel is small because the
threshold opens at 1093 MeV and its coupling is weaker than for $KK$
and $\eta \pi$.
However, its effect is not negligible.

\begin {table} [htb]
\begin {center}
\begin {tabular}{cc}
Sheet & Pole $M - i\Gamma /2$ (MeV) \\
\hline
$+-+$ & $(989.1 \pm 1.0 \pm 3.0) - i(40.1 \pm 1.9 \pm 2.7)$   \\
$+--$ & $(997.4 \pm 1.0 \pm 5.0) - i(46.3 \pm 1.9 \pm 4.2)$  \\
$+++$ & $(920  \pm 3  \pm 20) - i(93 \pm 5 \pm 20)$  \\\hline
\end {tabular}
\caption {Pole positions on sheets labelled by signs of $i$ in
channels $\eta \pi$, $KK$ and $\eta '\pi$.
The first errors are statistical and the second systematic.}
\end {center}
\end {table}

In Table 5, statistical errors are assigned from corresponding errors
for fitted values of $M$ and $g^2$.
Systematic errors arise from
(i) neglect of mass differences between the three $KK$
charge combinations,  (ii) uncertainty about form factors, (iii)
isospin mixing with $f_0(980)$.
The first of these is estimated from half the spread of $KK$ mass
differences.
The second is estimated from errors in the exponent of the form factor
for the $KK$ channel: $\alpha = 2.0 \pm 0.5$ GeV$^{-2}$.
This estimate is  obtained from experience in fitting many sets of
Crystal Barrel data for wider resonances, where the form factor has a
stronger effect.
The third error due to isospin violation is unknown at present.
However, one expects isospin mixing to produce
effects small compared with width differences of $f_0(980)$ and
$a_0(980)$; the second-sheet pole for $f_0(980)$ has $\Gamma /2 = 17
\pm 4$ MeV from current BES II data \cite {WPP}.
Systematic effects of the $\eta '\pi$ channel are hard to estimate
without data for that channel.
They have been estimated as half the difference between the first and
second entries of Table 5.

The position of the pole with $+++$ signs
is further from the physical region than the second-sheet pole.
It has changed greatly from earlier fits using Breit-Wigner amplitudes
without form factors or Adler zeros.
In that earlier work, this pole, commonly called the third-sheet pole,
lay close to $1040- i83$ MeV.
The large change in its mass arises from sensitivity to form factors.
In this respect, Crystal Barrel data fitted here, although they
determine $M$ and $g^2$ with modest errors, do not give accurate
information on the precise line-shape in $\eta \pi$.
This is because of uncertainties in interferences with the $\pi \pi$
S-wave and open questions about exactly how to parametrise it.
Forthcoming Kloe data should improve this situation substantially.
The line-shape in those data measures directly the form factor
for coupling of $\eta \pi$ to the resonance.
The systematic error assigned in Table 5 assumes the form factor has
exponent $\alpha = 2.0\pm 0.5$ GeV$^{-2}$, corresponding to a Gaussian
source with a root mean square radius $0.68 \pm 0.08$ fm.

Positions of the poles in the
sheets $++-$ and $+--$ corresponding to the $\eta '\pi$ channel will
not be given because of the absence of data for that channel.

\section {Conclusions}
Experimental conclusions are straightforward.
Firstly, dispersive corrections and the $s$-dependence of amplitudes
play a major role for both $a_0(1450)$ and $a_0(980)$.
The fit to $\bar pp \to \eta \pi ^0\pi ^0$ is decisive in settling
parameters of $a_0(1450)$.
Without these data, the width fitted to $\omega \pi ^+\pi ^-\pi ^0$
data alone is unreasonably small $\sim 110$ MeV.
With the inclusion of decays of $a_0(1450) \to a_0(980)\sigma$,
there is an excellent fit to both sets of data, giving the fitted
parameters of Table 2.
The $\chi^2$ of both fits improve with the inclusion of the
$s$-dependence of the $a_0(1450)$ amplitude.
More importantly, the fit stabilises in a narrow range of parameters
and the fitted $a_0(1450)$ signal in $\eta \pi ^0\pi ^0$ increases by
a factor 1.8; this is a clear indication of a better fit to
the line-shape.
The overall conclusion is that $a_0(1450)$ decays weakly to
$\eta \pi$, $KK$ and $\eta '\pi$ and dominantly to $\omega \rho$
and $a_0(980)\sigma$.
It is unfortunate that data on $\bar pp \to K^0_L K^\pm \pi ^\mp$
do not constrain the ratios $g^2_{KK}/g^2_{\eta \pi}$
for either $a_0(1450)$ or $a_0(980)$ tightly.
The data are consistent with the SU(3) predictions.

For $a_0(980)$, there is quite good agreement between Crystal Barrel
$\eta \pi ^0\pi ^0$ data and Kloe data.
Each have their merits.
A limitation of present data is the mass resolution of
Crystal Barrel near the $KK$ threshold, $\pm 9.5$ MeV.
If progress is to be made on isospin mixing between $a_0(980)$ and
$f_0(980)$, a mass resolution better than 0.5 MeV seems desirable,
i.e. 10\% of the spread of $KK$ masses.

An incidental result is a better understanding of the dominant
$J^{PC} = 0^{-+}$ signal in $\bar pp \to (\omega \rho )\pi$.
A plausible interpretation of the data is presented in terms of a
cusp at the $\omega \rho$ threshold and a radial recurrence
of $\pi (1300)$ close to 1730 MeV.
However, more precise data are needed to clarify this result in
$\omega \rho$, $\rho \pi$ and $[\sigma \pi ]_{L=1}$ channels.
The Compass experiment could be a good source of such data.
Obviously a search for the missing $a_0$'s expected
at higher mass is sorely needed.

As regards the interpretation of $a_0(1430)$, it seems likely to
be dominantly an $n\bar n$ state.
Black, Fariborz and Schechter have pointed out that
$a_0(980)$ may have a radial excitation in the general mass range
of $a_0(1430)$ \cite {Fariborz}.
Such a radial excitation would almost inevitably mix with the
expected $n\bar n$ state.
The radial excitation would respond to long range meson-meson
interactions, and therefore to the $a_0(980)\sigma$ and $\omega \rho$
thresholds.
These are likely to be responsible for pushing the mass of $a_0(1450)$
up to that of $K_0(1430)$.

One of the essential ingredients in fitting all the data is the
$\pi \pi$ S-wave (and $K\pi$).
Empirically it is necessary to parametrise it with
the 2-component form of Eq. (8).
This gives considerably flexibility to the numerator $N(s)$ of the
amplitude, even though the denominator $D(s)$ is accurately known.
Guidance from theory on the way the $\sigma$ and $\kappa$ couple in
production reactions would be very helpful to experimentalists.

\end {document}